\documentclass[reqno]{amsart}
\usepackage{amsmath,amssymb}
\usepackage{tikz,pgf}
\usetikzlibrary{arrows,decorations.pathmorphing,backgrounds,positioning,fit,petri}
\usetikzlibrary{calc,intersections,through}
\usepackage{amsmath,amssymb,cite} 
\usepackage{graphics,epsfig,color}
\usepackage[mathscr]{euscript}
\theoremstyle{plain}
\newtheorem{theorem}{Theorem}[section]

\newtheorem{lemma}[theorem]{Lemma}

\theoremstyle{definition}

\theoremstyle{remark}
\newtheorem{remark}[theorem]{Remark}

\numberwithin{equation}{section}
\numberwithin{theorem}{section}

\renewcommand{\epsilon}{\varepsilon}
\renewcommand{\tilde}{\widetilde}

\renewcommand{\div}{\mathop{\rm div}\nolimits}

\definecolor{light}{gray}{.9}

\begin{document}

%
%
%
%
%
%
%
%
%

\title[The alphabet model]{The energy of the alphabet model }

\author[D.\ Gabrielli]{Davide Gabrielli}
\address{ DISIM, University of L'Aquila
  \hfill\break\indent
  Via Vetoio,   67100 Coppito, L'Aquila, Italy}
\email{gabriell@univaq.it}


\author[F. Roncari]{Fabio Roncari}
\address{DISIM, University of L'Aquila
  \hfill\break\indent
  Via Vetoio,   67100 Coppito, L'Aquila, Italy}
\email{fabio.roncari@gmail.com}



\date{January 1, 2004}

\begin{abstract}
We call \emph{Alphabet model} a generalization to N types of particles of the classic ABC model.
We have particles of different types stochastically evolving on a one dimensional lattice with an exchange dynamics.
The rates of exchange are local but under suitable conditions the dynamics is reversible with a Gibbsian like invariant measure with
long range interactions. We discuss geometrically the conditions of reversibility on a ring that correspond to a gradient condition on the graph of configurations or equivalently to a divergence free condition on a graph structure associated to the types of particles. We show that much of the information on the interactions between particles can be encoded in  associated \emph{Tournaments} that are a special class of oriented directed graphs. In particular we show that the interactions of reversible models are corresponding to strongly connected tournaments. The possible minimizers of the energies are in correspondence with the Hamiltonian cycles of the tournaments. We can then determine how many and which are the possible minimizers of the energy looking at the structure of the associated tournament. As a byproduct we obtain a probabilistic proof of a classic Theorem of Camion \cite{Camion} on the existence of Hamiltonian cycles for strongly connected tournaments. Using these results we obtain in the case of an equal number of k types of particles new representations of the Hamiltonians in terms of translation invariant $k$-body long range interactions. We show that when $k=3,4$ the minimizer of the energy is always unique up to translations. Starting from the case $k=5$ it is possible to have more than one minimizer. In particular it is possible to have minimizers for which particles of the same type are not joined together in single clusters.
\end{abstract}

\maketitle
\section{Introduction}
The ABC model is a simple but interesting stochastic lattice gas having particles of three types, say A, B and C, evolving on a lattice. It has been introduced and studied on a one dimensional ring in \cite{EKKM1,EKKM2}. The peculiar feature of the model is that, while the rates of transition are local functions of the configuration, in the case of an equal number of particles of each type the model is reversible with respect to an invariant measure written in terms of an Hamiltonian with long range interactions. Due to the presence of long range interactions, even if the model is one dimensional, it exhibits a phase transition. Since its introduction the model and its generalizations have been quite intensively studied, see for example \cite{ACLMMS, BLS, BCP, BDGJL, BD, CDE, FF1, FF2}.

We study a generalization of this model with up to $N$ different types of particles where $N$ is the size of the lattice. The model with more than 3 types of particles has been already studied in \cite{EKKM2, FF1, FF2}. We call the case with $N$ different type of particle the {\emph alphabet model}. Particles that are at the two extremes of a bond of the lattice can exchange their position and the rate at which this happens depends only on the configuration restricted to the bond and not on the structure of the configuration on different sites. We mainly consider the model on the one dimensional ring but we will discuss also in some aspects the model on a one dimensional interval.

We first discuss the conditions to have reversibility. This result has been already discussed in \cite{EKKM2, FF1, FF2}. Our approach has a geometric flavor and moreover it gives a new interpretation to these conditions. Reversibility coincides with a divergence free condition on an abstract graph whose vertices are labeled by the types of particles evolving. The use of this graphical representation is very useful to understand the dynamics of the system. We show indeed how it is possible to encode much of the information about the interactions among the particles with a {\emph Tournament} having vertices labeled by the type of particles. A tournament is a complete oriented graph.

The energy of the alphabet model is naturally written in terms of long range two body interactions. We show however that in the case of $k$ types of particles evolving on a ring with $N$ sites and such that there are $\frac Nk$ particles of each type the energy can be naturally written also in terms of long range k body interactions. The advantage of this new representation is on the fact that it is manifestly translational invariant, it has a natural combinatorial interpretation and above all it is very useful for computations.

We prove that the reversible alphabet models are in correspondence with strongly connected tournaments. We argue also that the possible minimizers of the energy are in correspondence with the Hamiltonian cycles of the associated tournament. We obtain in this way a probabilistic proof of the classic Camion \cite{Camion} Theorem on the existence of Hamiltonian cycles for strongly connected tournaments.

Tournaments with a small number of vertices have been classified \cite{Moon} and using this classification we can easily analyze the possible reversible alphabet dynamics with a few types of particles. We obtain for example that with 3 types of particles essentially the classic ABC model is the unique possible dynamics. Quite surprisingly this is true also with 4 types of particles. In this case there is a 3 parameter family of reversible models but changing names to the particles all of them will have the same topological structure of the interaction so that in a sense they will be all equivalent. This follows by the fact that there is only 1 unlabelled strongly connected tournament with 4 vertices. We obtain also that the minimizer of the energy is always unique since this tournament has a unique Hamiltonian cycle.
The situation changes with 5 types of particles. In this case several strongly connected tournaments exist and some of them have more than one Hamiltonian cycle. In this way we obtain models for which it is possible to have more than one minimizer of the energy. Using the 5 body representation of the energy for a model with $\frac N5$ particles of each type we show that it is possible for example to have minimizers of the energy where the particles of each type are not joined all together in a single cluster.

\smallskip
The structure of the paper is the following. In section 2 we briefly discuss the graph theoretical tools that we are using in the paper. In section 3 we define the models. In section 4 we summarize the main results of the paper. In section 5 we study the conditions to have reversibility. In section 6 we compute the energy. In section 7 we show how it is possible to encode information on the interactions among the particles on a Tournament and obtain a proof of Camion Theorem. In section 8 we discuss the minimizers of the energy for 2,3,4 and 5 types of particles.

\section{Graph theoretical tools}
In this section we briefly review some notions of graph theory that we will use in the paper. An oriented graph is a pair $(V,E)$ where $V$ is the collection of vertices while $E$ is the collection of ordered edges that are ordered pairs $(x,y)$ of elements of $V$. We consider graphs without loops, i.e. edges of the form $(x,x)$. To an oriented graph $(V,E)$ we associate the un-oriented graph $(V,\mathcal E)$ where an un-oriented edge $\{x,y\}$ belongs to $\mathcal E$ if and only if at least one between $(x,y)$ and $(y,x)$ belongs to $E$. An oriented path is a sequence $(x_0,\dots ,x_n)$ such that $(x_i,x_{i+1})\in E$ for any $i$. If we require just $\{x_i,x_{i+1}\}\in \mathcal E$ for any $i$ then we have an un-oriented path. A self avoiding path with just the initial and the final vertices coinciding is a cycle. Two cycles crossing the same oriented edges and differing just for the starting points are naturally identified. An oriented graph is strong or strongly connected if for any pair of vertices there exists an oriented path connecting them. An un-oriented  graph is connected if there exists an un-oriented path connecting any pair of vertices.

\smallskip

It will be very useful to represent graphically some properties in terms of graphs. There will be 3 different types of graphs depending on the set of vertices.

\noindent \emph{The physical graph}: this is the one dimensional lattice, a ring or an interval, that represents the physical space where the particles are moving. Vertices of this graph are usually denoted by $x,y,z$.

\noindent \emph{The configuration space}: in this case vertices are labeled by the configurations of particles $\eta$ while the edges are in correspondence of the pairs of configurations $(\eta,\eta')$ for which it is possible to transform $\eta$ into $\eta'$ with a single transition of the Markov dynamics.

\noindent \emph{The graphs of types}: we will introduce graphs having vertices labeled by the different types of evolving particles. This means that if there are $k$ different types of particles then these graphs will have $k$ vertices. In the case for example of the ABC model a graph of this type will have 3 vertices. Vertices for graphs of this kind will be labeled like $i,j,l$.

\subsection{Flows and discrete vector fields}

 Given an oriented graph $(V,E)$, a flow is a map $Q:E\to \mathbb R^+$ that associates to the edge $(x,y)$ the amount of flowing mass
 $Q(x,y)$. The divergence of a flow $Q$ at $x\in V$ is defined by
 \begin{equation}
 \div Q(x):=\sum_{y:(x,y)\in E}Q(x,y)-\sum_{y:(y,x)\in E}Q(y,x)\,,
 \end{equation}
 and represents the amount of mass that is flowing outside $x$ minus the amount of mass that is flowing into $x$.

 For oriented graphs such that if $(x,y)\in E$ then also $(y,x)\in E$ we define a \emph{discrete vector field} as a map $\phi:E\to \mathbb R$ that is antisymmetric i.e. satisfies $\phi(x,y)=-\phi(y,x)$. The divergence of a discrete vector field at $x\in V$ is defined by
\begin{equation}\label{divdisc}
\div \phi(x):=\sum_{y\,:\,(x,y)\in E}\phi(x,y)\,.
\end{equation}
A discrete vector field $\phi$ is called of gradient type if there exists a function $f$ such that $\phi(x,y)=f(y)-f(x)$.
Gradient vector fields can be characterized by the condition $\sum_i\phi(x_i,x_{i+1})=0$ on any cycle.

A classic result on discrete vector fields is the Hodge decomposition \cite{Biggs}. Let us call $\Lambda^1$ the vector space of discrete vector fields on a given oriented graph $(V,E)$. The dimension of this vector space is $\frac{|E|}{2}$. We endow $\Lambda^1$ with the scalar product
\begin{equation}\label{sc}
\langle \phi,\psi\rangle:=\frac 12\sum_{(x,y)\in E}\phi(x,y)\psi(x,y)\,, \qquad \phi,\psi \in \Lambda^1\,.
\end{equation}
The Hodge decomposition states that $\Lambda^1$ can be orthogonally decomposed like
\begin{equation}\label{ancorahodge}
  \Lambda^1=\Lambda^1_g\oplus\Lambda^1_d\,,
\end{equation}
where $\Lambda^1_g$ is the subspace of gradient vector fields while $\Lambda^1_d$ is the subspace of divergence free discrete vector fields; the orthogonality is with respect to the scalar product \eqref{sc}. The dimension of $\Lambda^1_g$ is $|V|-1$ while the dimension of $\Lambda^1_d$ is $\frac{|E|}{2}-|V|+1$. A basis for $\Lambda^1_d$ can be constructed considering a collection of independent cycles of $(V,\mathcal E)$. A cycle of the basis is obtained starting from a spanning tree and adding one edge (see \cite{Biggs} for more details).

\smallskip

\noindent The following is essentially Lemma 4.3 of \cite{IHP} in the finite case and will be useful in the following.
\begin{lemma}\label{IHPqui}
A finite oriented graph $(V, E)$ is strongly connected if and only if $(V,\mathcal E)$ is connected and there exists a flow $Q$ with $Q(x,y)>0$ for any $(x,y)\in  E$ and $\div Q=0$.
\label{tommy}
\end{lemma}
For the proof we refer to \cite{IHP}.

\subsection{Tournaments}
The classic book on tournaments is \cite{Moon} to which we refer for all the details. Several results on tournaments can be found also in \cite{D}.
A tournament is an oriented graph $(V,E)$ such that if $i,j$ are two vertices then exactly one between $(i,j)$ and $(j,i)$ is an element of $E$. This means that the corresponding un-oriented graph $(V,\mathcal E)$ is the complete graph on $|V|$ vertices; moreover for each un-oriented edge of this complete graph exactly one between the two possible orientations is an oriented edge of the original tournament. A simple way of constructing a tournament is then to start with a complete un-oriented graph with $|V|$ vertices and chose one arbitrary orientation for each un-oriented edge.

There are two basic notions related to tournaments. A tournament is called \emph{strong} if it is strongly connected as an oriented graph.
A tournament is called \emph{reducible} when it is possible to divide the set of vertices into two nonempty sets $A$ and $B$ such that the edges having one extreme vertex in $A$ and the other in $B$ are all oriented from the vertex in $A$ to the one in $B$. A tournament not reducible is called \emph{irreducible}. These two notions indeed coincide since a tournament is strong if and only if it is irreducible (see Theorem 2 of \cite{Moon})

An oriented graph is called \emph{Hamiltonian} if there exists a cycle visiting once all the vertices and starting and ending in the same vertex. Such a cycle is called an Hamiltonian cycle. More precisely an oriented graph is Hamiltonian if there exists an ordering
$\left(i_1,\dots i_{|V|}\right)$ of the vertices of the graph such that $(i_n,i_{n+1})\in E$ for $n=1,\dots ,|V|$ (the sum in the indices is modulo $|V|$ so that $i_{|V|+1}=i_1$). If instead we do not require that $(i_{|V|},i_1)\in E$ then we obtain an Hamiltonian path. A graph that contains an Hamiltonian path is called \emph{traceable}.

A basic fact about tournaments (Theorem 1.4.5 of \cite{D}) is that every tournament is traceable.
A classic result on tournaments firstly obtained by Camion \cite{Camion} is that a tournament is strong if and only if it is Hamiltonian (see also Theorem 1.5.1 in \cite{D}).

Two tournaments are \emph{isomorphic} if there exists a one-to-one
correspondence between their nodes that is compatible with the orientation of the edges.
An equivalence class of isomorphic tournaments can be represented by a tournament with un-labeled vertices. The different tournament belonging to the class can be obtained giving different labeling to the vertices. A classification of non isomorphic tournaments with few vertices is available (see the appendix of \cite{Moon}).

\section{The models}\label{mod}

We consider exclusion processes with several types of particles evolving with an exchange dynamics. We consider mainly three
different frameworks that are the following.

\subsection{The ABC model}
We first recall the classic ABC model on a one dimensional periodic lattice of length $N$. On the ring there are three types of particles called respectively particles of type A, particles of type B and particles of type C. Every vertex of the lattice is occupied by exactly one particle. This means that there are no empty sites and no sites with more than one particle.
A configuration of particles is encoded by $\eta=(\eta(x))_{x\in \mathbb Z_N}=(\eta_A(x),\eta_B(x),\eta_C(x))_{x\in \mathbb Z_N}$, where $x\in \mathbb Z_N:=\mathbb Z/N\mathbb Z$
is a vertex of the lattice. The binary variables $\eta_A(x)$, $\eta_B(x)$ and $\eta_C(x)$ can assume just the values $0$ and $1$. If on a site $x$ there is a particle of type $Y$, where $Y$ can be A or B or C, then $\eta_Y(x)=1$, otherwise we have $\eta_Y(x)=0$. We draw the periodic lattice $\mathbb Z_N$ as a ring on which the anticlockwise orientation corresponds to the direction going from $x$ to $x+1$ (the sum is modulo $N$).

The ABC stochastic dynamics is a dynamics that conserves the total number of particles of the three types and is defined as follows. The generator can be easily written as
\begin{equation}\label{gendav}
\mathcal L_N f(\eta)=\sum_{x\in \mathbb Z_N}c_{x,x+1}(\eta)\left[f(\eta^{x,x+1})-f(\eta)\right]
\end{equation}
where $\eta^{x,x+1}$ is the configuration obtained from the configuration $\eta$ exchanging the values at sites $x$ and $x+1$
\begin{equation}\label{exchange-dav}
\eta^{x,x+1}(y)=\left\{
\begin{array}{ll}
\eta(x+1) & \textrm{if}\ y=x\,, \\
\eta(x) & \textrm{if}\ y=x+1\,, \\
\eta(y) & \textrm{otherwise}\,.
\end{array}
\right.
\end{equation}
The rates of exchange are defined as
\begin{align}
c_{x,x+1}(\eta)&:= q\left[\eta_A(x)\eta_B(x+1)+\eta_B(x)\eta_C(x+1)+\eta_C(x)\eta_A(x+1)\right]\nonumber \\
&+
\left[\eta_B(x)\eta_A(x+1)+\eta_C(x)\eta_B(x+1)+\eta_A(x)\eta_C(x+1)\right]\,.\label{ratesdav}
\end{align}
The sums inside the squared parenthesis can assume only the values $0$ and $1$ and we can summarize formula \eqref{ratesdav} just saying
that if we have in two neighboring sites $x,x+1$ a configuration of the type AB or BC or CA then the configuration
transforms with rate $q$ into respectively BA, CB and AC; if instead we have in two neighboring sites $x,x+1$ a configuration of the type BA or CB or AC then the configuration
transforms with rate $1$ into respectively AB, BC and CA. Typically $q$ is a positive number smaller than one. This means that locally on each single bond $(x,x+1)$ the configurations of the type AB or BC or CA are stable and change slower than the unstable configurations that are of the form BA or AC or CB.

The dynamics is well defined for any number of particles of the different type. The case when the numbers of particles of type A, B and C are all equal to $\frac N3$, that is $\sum_x\eta_A(x)=\sum_x \eta_B(x)=\sum_x\eta_C(x)=\frac N3$, is a special case since the model is reversible with respect to a Gibbsian invariant measure
\begin{equation}\label{mis-dav}
\mu_N=\frac{e^{-H_N(\eta)}}{ Z_N}
\end{equation}
where $Z_N$ is a normalization constant.
The Hamiltonian $H_N$ has several equivalent representations in terms of long range two body interactions. The most used are
\begin{equation}\label{prima-ham-dav}
(\log q)\sum_{x=1}^{N-1}\sum_{y= x+1}^N \left[\eta_C(x)\eta_B(y)+\eta_B(x)\eta_A(y)+\eta_A(x)\eta_C(y)\right]
\end{equation}
and
\begin{equation}\label{ham2-dav}
\frac {(\log q)}{N}\sum_{x=1}^N\sum_{y=1}^{N-1}y\left[\eta_A(x)\eta_B(x+y)+\eta_B(x)\eta_C(x+y)+\eta_C(x)\eta_A(x+y)\right]\,.
\end{equation}
The Hamiltonians in \eqref{prima-ham-dav} and \eqref{ham2-dav} are not equal but differ by a constant factor. This constant factor is irrelevant since they  induce  the same probability measure.

\subsection{The k-types case}

We now generalize the notation considering $k$ different types of particles evolving on a one dimensional periodic lattice of length $N$. The dynamics can be defined for any $k\leq N$ and for any number of particles of each type but we will discuss results
just for a special situation. This is when $\frac Nk$ is an integer number and there are exactly $\frac Nk$ particles of each type.

On the ring there are $k$ types of particles called respectively particles of type $1,2,\dots , k$. Every vertex of the lattice is occupied by exactly one particle. This means that there are no empty sites and no sites with more than one particle.
A configuration of particles is encoded by $\eta=(\eta(x))_{x\in \mathbb Z_N}=(\eta_1(x),\dots ,\eta_k(x))_{x\in \mathbb Z_N}$. The binary variables $\eta_j(x)$ can assume just the values $0$ and $1$. If on a site $x$ there is a particle of type $j$, then $\eta_j(x)=1$, otherwise we have $\eta_j(x)=0$.

The stochastic dynamics is defined as follows (see also \cite{EKKM2, FF1, FF2}).
The generator is still  \eqref{gendav} and $\eta^{x,x+1}$ is again the configuration obtained from the configuration $\eta$ exchanging the values at sites $x$ and $x+1$ like in \eqref{exchange-dav}.
The rates of exchange $c_{x,x+1}(\eta)$ are translational covariant (this means $c_{x,x+1}(\eta)=c_{x+z,x+z+1}(\tau_z\eta)$ where $\tau_z$ denotes translation by $z$) and depends only on the occupation variables $\eta(x)$ and $\eta(x+1)$ and not on the structure of the configuration of particles outside these two sites. More precisely we consider a positive function $c:\{1,\dots ,k\}\times \{1,\dots ,k\}\to \mathbb R^+$ and define the exchange rates in \eqref{gendav} as
\begin{equation}\label{exrra}
c_{x,x+1}(\eta):=c(\eta(x),\eta(x+1))\,.
\end{equation}
This means that when site $x$ is occupied by a particle of type $i$ while site $x+1$ is occupied by a particle of type $j$
then $c(i,j)$ is the corresponding rate of exchange. Note that in general $c(i,j)\neq c(j,i)$; for example in the ABC model we have $c(A,B)=q\neq c(B,A)=1$.

Sometimes it will be useful to encode the configuration of particles with $\tilde\eta=\left(\tilde\eta(x)\right)_{x\in \mathbb Z_N}\in\{1,\dots ,k\}^{\mathbb Z_N}$ defined by setting $\tilde\eta(x)=i$ if at site $x$ there is a particle of type $i$.

\subsection{The alphabet model}
The cases when $k=N$ play an important role in our analysis. When the number of different types of particles coincides with the number of sites we say that we have an alphabet model.
The generator is still \eqref{gendav} and the exchange rates are given by \eqref{exrra} where $c:\{1,\dots ,N\}\times \{1,\dots ,N\}\to \mathbb R^+$. In this case we use also the notation $x_\eta(i)$ to denote the element of the ring where is located the particle $i$ in the configuration $\eta$. When the configuration $\eta$ is clear we do not write explicitly the dependence on $\eta$. Given $x,y\in \mathbb Z_N$ we denote by $[x,y]$ the interval of the ring containing the sites that are encountered starting from $x$ and moving anticlockwise up to reach $y$.

\section{Summary of the main results}
All the results of the paper are completely rigorous and could be summarized into a collection of Theorems. We prefer however, to have more flexibility in the presentation, to illustrate them without this rigid constraint. For the readers convenience we summarize in this section all the results considering separately the different sections.

It is useful to define
\begin{equation}\label{gammai}
\gamma(i,j):=\log\frac{c(i,j)}{c(j,i)}\,.
\end{equation}
Since it satisfies $\gamma(i,j)=-\gamma(j,i)$ it is a discrete vector field on the graph having particle types as vertices and all the possible edges. Several of the results depend just on the discrete vector field $\gamma$.

\subsection{Reversibility}
In Section 5 we characterize completely the reversibility of the alphabet model on the ring and on an interval. On the ring we obtain that the alphabet model is reversible if and only if $\div \gamma=0$, i.e. the vector field $\gamma$ is divergence free on the graph having particle types as vertices and all the possible edges. On an interval we have instead that any alphabet model is reversible. We use a geometric approach based on the Kolmogorov criterion for reversibility.
Some of the results were already obtained in \cite{EKKM2, FF1, FF2}.
We give a different proof with a more direct interpretation.
The reversibility of the k-types models can be obtained as a special case.

\subsection{The energy}
In Section 6 we study the energy associated to the invariant measure of the models that has long range interactions
even if the dynamics has local rules. In the case of a reversible alphabet model on the ring we give two different representations that are formulas \eqref{io} and \eqref{copa}. Formula \eqref{copa} holds also for any alphabet model on an interval.

\smallskip
In the case of the ABC model on the ring with $\frac N3$ particles of each type we show that there is a very natural way of writing the invariant measure as a Gibbs measure \eqref{mis-dav} with an energy having long range translation invariant 3-body interactions
\begin{equation}\label{3bodyi}
H_N(\eta)=\sum_{x,y,z\in \mathbb Z_N}J(x,y,z)\eta_A(x)\eta_B(y)\eta_C(z)\,.
\end{equation}
The interactions $J$ are of topological type and we have that $J(x,y,z)=\frac{3\log q}{N}$ if going anticlockwise on the ring we meet $z$ before $y$ and  $J(x,y,z)=0$ otherwise. Correspondingly there is a similar representation
of the large deviation rate functional for the invariant measure in the weakly asymmetric regime \eqref{LDP-rate-dav}.

\smallskip
Also in the case of a reversible model with $k$ types of particles on the ring and having $\frac Nk$ particles of each type we can write the invariant measure as a Gibbs measure \eqref{mis-dav} with an energy having translation invariant long range k-body interactions
\begin{equation}\label{kbodyi}
H_N(\eta)=\sum_{x_1,\dots,x_k}J(x_1,\dots,x_k)\eta_1(x_1)\dots\eta_k(x_k)\,.
\end{equation}
The interactions $J$ have also a topological structure that is described in the corresponding section \ref{ktypesec}.

\subsection{The graph of the interaction}
In Section 7 we associate to each alphabet model a tournament that encodes the topological structure of the interaction among the particles. We show that tournaments associated to reversible alphabet models are necessarily strongly connected. Conversely given a strongly connected tournament it is possible to construct a reversible alphabet model with associated the given tournament.

We show that on a ring the minimizers of the energy for a reversible alphabet model are in correspondence with the Hamiltonian cycles of the tournament while on an interval with the Hamiltonian paths. Using the fact that the energy has at least one minimizer we
obtain that every strongly connected tournament has at least one Hamiltonian cycle and that every tournament has at least one Hamiltonian path (we use respectively a model on the ring and on the interval). These are classic statements in the theory of tournaments \cite{Moon,D}, the first one is due to Camion \cite{Camion}.

\subsection{Minimizers}
In Section 8, using the classification of tournaments with a small number of sites \cite{Moon}, we show that all the reversible alphabet models on the ring with $N=3,4$ have, up to a permutation on the labels of the particles, the same associated tournament that moreover has an unique Hamiltonian cycle. This implies that the minimizer of the energy is always unique (up to rotations). Starting from $N=5$ we have different possible tournaments and models with more than one single minimizer.

\smallskip

The study of the minimizers of the energy can be done also in the case of k types of particles with $\frac{N}{k}$ particles of each type. Again on the ring we obtain for $k=3,4$ that the minimizer is again unique up to rotations and moreover all the particles of the same type are joined together in single clusters. When $k=5$ instead we have several possible situations and in particular we can have more than one minimizers and minimizers on which particles of the same type belong to different clusters.

\section{Reversibility}\label{rever}
We discuss in detail the alphabet case. As illustrated at the end, the k-type case can be obtained as a special case

\subsection{The alphabet model on a ring}
The conditions to have reversibility in this case have been already obtained in \cite{EKKM2, FF1, FF2}. We give a different geometric approach and an interpretation of the conditions on the graph of types. For a continuous time Markov chain having transition rate of jump from configuration $\eta$ to $\eta'$ given by $r(\eta,\eta')$, reversibility with respect to a Gibbsian probability measure like \eqref{mis-dav} coincides with the validity of the detailed balance condition
\begin{equation}\label{db}
e^{-H_N(\eta)}r(\eta,\eta')=e^{-H_N(\eta')}r(\eta',\eta)\,.
\end{equation}
Condition \eqref{db} is equivalent to impose that $\log\frac{r(\eta,\eta')}{r(\eta',\eta)}$ is a discrete vector field of gradient type on the graph of the configuration space. The corresponding function is $-H_N$. The Kolmogorov criterion for reversibility corresponds to impose that
\begin{equation}\label{kol}
\sum_i\log\frac{r(\eta^{(i)},\eta^{(i+1)})}{r(\eta^{(i+1)},\eta^{(i)})}=0
\end{equation}
along all possible cycles $(\eta^{(0)},\eta^{(1)},\dots ,\eta^{(n)},\eta^{(0)})$ on the configuration space that is exactly the condition that characterizes gradient discrete vector fields.

Let
\begin{equation}\label{gamma}
\gamma(i,j):=\log\frac{c(i,j)}{c(j,i)}\,.
\end{equation}
 Since it satisfies $\gamma(i,j)=-\gamma(j,i)$ it is a discrete vector field on the graph having particle types as vertices and all the possible edges. We will show that on this graph the reversibility of the alphabet model is exactly equivalent to impose that $\gamma$ is a divergence free discrete vector field, i.e. that
\begin{equation}\label{rev}
\sum_{j\neq i}\gamma(i,j)=\div\gamma(i)=0\,, \qquad i=1,2,\dots ,N\,.
\end{equation}

To show this statement let us consider a generic cycle $\mathcal C:=(\eta^{(0)},\eta^{(1)},\dots$ $ ,\eta^{(n)},\eta^{(0)})$ on the configuration space. Since the starting and ending configurations coincide with $\eta^{(0)}$ then the corresponding trajectory of every particle will be also a closed cycle on the one dimensional physical lattice $\mathbb Z_N$. In particular to any particle there will be associated a winding number $W$. In particular $W(i)$ is an integer number indicating the number of times particle $i$ is turning around the ring in its cyclic trajectory. This number is positive or negative depending on whether the net rotation is clockwise or anticlockwise.  Let us also call $\mathcal N_{i,j}$ the number of times that in the cycle $\mathcal C$ we exchange the position of the $i$ particle with the $j$ particle and the exchange is such that the $i$ particle is moving anticlockwise while the $j$ particle is moving clockwise. Clearly $\mathcal N_{j,i}$ represents the number of exchange of positions between particles $i$ and $j$ but such that the particles are moving in the opposite directions. A basic topological fact says that their difference is fixed by the winding numbers since we have the following basic identity
\begin{equation}\label{topo1}
\mathcal N_{i,j}-\mathcal N_{j,i}=W(i)-W(j)\,.
\end{equation}
The validity of this identity is illustrated in Figure \ref{fig:windingNumbers}. We represent on the
horizontal axis periodically the one dimensional ring. On the vertical axis there is the natural parameter describing the evolution on the cycle $\mathcal C$. By periodicity the trajectory of each particle can be represented by infinitely many curves just horizontally shifted. We draw all of them for example for the particle $i$ (green lines) and just one for the particle $j$ (blue line). The number $\mathcal N_{i,j}$ is obtained just counting the number of crossings of the blue line with a green line such that the blue line is going from the right of the green line to the left. Recall that the orientation of the curves is in accordance with the parameter of evolution on the cycle $\mathcal C$ so that it is represented from the bottom toward the top. The number $\mathcal N_{j,i}$ is instead obtained just counting the number of crossings of the blue line with a green line such that the blue line is going from the left of the green line to the right. Fixed a green line the number of crossing from right to left minus the number of crossing from left to right depends only on the starting and ending points of the two curves and can be only $-1,0,+1$. In particular it is $\pm 1$ if the order of the final points of the blue and the green lines is different from the order of the corresponding initial points. It is instead zero if the same order is preserved. From these arguments it is now easy to get \eqref{topo1}.
\begin{figure}
\setlength{\unitlength}{0.5cm}
\begin{tikzpicture}[scale=0.5]
\draw (0,0) to (20,0);
\draw (0,10) to (20,10);
\draw[dotted] (5,0) to (5,10);
\draw[dotted] (10,0) to (10,10);
\draw[dotted] (15,0) to (15,10);
\draw[dotted] (20,0) to (20,10);
\draw[dotted] (0,0) to (0,10);
\draw[thick,color=blue!80,domain=0:4] plot ({-(\x^2)/2+2*\x+4},\x);
\node [shape=circle,fill=black!20,draw] (id1) at    (1,0) {i};
\node [shape=circle,fill=black!20,draw] (jd1) at    (4,0) {j};
\node [shape=circle,fill=black!20,draw] (iu1) at    (1,10) {i};
\node [shape=circle,fill=black!20,draw] (ju1) at    (4,10) {j};
\node [shape=circle,fill=black!20,draw] (id2) at    (6,0) {i};
\node [shape=circle,fill=black!20,draw] (jd2) at    (9,0) {j};
\node [shape=circle,fill=black!20,draw] (iu2) at    (6,10) {i};
\node [shape=circle,fill=black!20,draw] (ju2) at    (9,10) {j};
\node [shape=circle,fill=black!20,draw] (id3) at    (11,0) {i};
\node [shape=circle,fill=black!20,draw] (jd3) at    (14,0) {j};
\node [shape=circle,fill=black!20,draw] (iu3) at    (11,10) {i};
\node [shape=circle,fill=black!20,draw] (ju3) at    (14,10) {j};
\node [shape=circle,fill=black!20,draw] (id4) at    (16,0) {i};
\node [shape=circle,fill=black!20,draw] (jd4) at    (19,0) {j};
\node [shape=circle,fill=black!20,draw] (iu4) at    (16,10) {i};
\node [shape=circle,fill=black!20,draw] (ju4) at    (19,10) {j};
\draw[thick,color=blue!80,domain=4:7] plot ({2*\x^2-18*\x+44},\x);
\draw[thick,color=blue!80,domain=16:18] plot (\x,{9*(\x-16)^2/20+(\x-16)/10+7});
\draw[thick,->,color=blue!80] (18,9) .. controls (+18.5,+9.6) and (18,9) ..  (ju4);
\draw[thick,->,color=green!50] (id1) .. controls (+5,+1) and (6,6) .. (iu1);
\draw[thick,->,color=green!50] (id2) .. controls (+10,+1) and (11,6) .. (iu2);
\draw[thick,->,color=green!50] (id3) .. controls (+15,+1) and (16,6) .. (iu3);
\draw[thick,->,color=green!50] (id4) .. controls (+20,+1) and (21,6) .. (iu4);

\end{tikzpicture}
\caption{Trajectories of particles i and j with winding numbers respectively $W(i)=0$ and $W(j)=3$.
On the horizontal axis we represent the physical one dimensional ring periodically on the real line.
On the vertical axis there is the natural parameter describing the evolution along the cycle $\mathcal C$. The trajectory of the particle $j$ is represented once (blue line).}
\label{fig:windingNumbers}
\end{figure}
Using the antisymmetry of $\gamma$, formula \eqref{topo1} and a discrete integration by parts, we obtain that the sum along the cycle $\mathcal C$ of $\log\frac{r(\eta,\eta')}{r(\eta',\eta)}$ is
\begin{align}
&\sum_l\log\frac{r(\eta^{(l)},\eta^{(l+1)})}{r(\eta^{(l+1)},\eta^{(l)})}=\sum_{i\neq j}\mathcal N_{i,j}\gamma(i,j)\nonumber  \\
&= \frac 12\sum_{i\neq j}\gamma(i,j)\left( \mathcal N_{i,j}-\mathcal N_{j,i}\right)=
\frac 12\sum_{i\neq j}\gamma(i,j)\left( W(i)-W(j)\right)\nonumber \\
&= \sum_i W(i)\left(\sum_{j\neq i}\gamma(i,j)\right)=\sum_i W(i)\div \gamma(i)\,.\label{itad}
\end{align}
To have reversibility we have that the right hand side of \eqref{itad} has to be zero for any possible choice of the winding numbers.
We observe that the winding numbers cannot be arbitrary integer numbers. This follows by the following argument. In a single exchange between two particles there is one particle that jumps to the right and one particle that jumps to the left. This implies that for any cycle $\mathcal C$ the total number of jumps of particles to the right minus the total number of jumps of particles to the left should be zero and this difference coincides with $N\sum_{i}W(i)$. We then obtained that for any cycle $\mathcal C$ in the configuration space we have
\begin{equation}\label{cococo}
\sum_iW(i)=0\,.
\end{equation}
Conversely, by a direct construction, it is possible to exhibit cycles on the space of configurations with associated any collection of integer winding numbers satisfying condition \eqref{cococo}. This is done for example considering particle 1 and moving it letting cross $W(1)$ times all the other particles (anticlockwise if $W(1)$ is positive and clockwise if negative). Then doing the same for particle $2$ and so on up to particle $N$.

The right hand side of \eqref{itad} is zero for any collection of winding numbers satisfying \eqref{cococo} if and only if
$$\div \gamma(i)=\lambda\,, \qquad \forall i$$
where $\lambda$ is an arbitrary real number. Since using the definition of discrete divergence we get that $\sum_i\div \gamma(i)=0$ we have that the only possible value is $\lambda=0$.
This gives condition \eqref{rev}.

Since the reversible alphabet models are in correspondence with the divergence free discrete vector field on the complete graph with $N$ vertices we know that the number of free parameters that we can use to parametrize the $\gamma$ of reversible models coincides with the dimension of $\Lambda^1_d$ that is $\frac{N^2-3N+2}{2}$.

Condition \eqref{rev} has the following simple interpretation. Consider any configuration and let particle $i$ move anticlockwise around the circle exchanging its position with all the other particles. The final configuration that we obtain in this way is a configuration coinciding with the initial one but with all the particles shifted by one unit in the clockwise direction. The sum of $\log\frac{r(\eta,\eta')}{r(\eta',\eta)}$ along this special path in the configuration space is exactly $\div \gamma(i)$. This means that the Kolmogorov criterion is satisfied as soon as it is satisfied for these N special elementary paths obtained letting one single particle go around the ring. If the model is reversible we obtain in this way that $H_N(\eta)=H_N(\tau_x\eta)$. Clearly this property can also be directly deduced by the translational covariance of the model. This means that if we are interested just in the energy we can then identify in an equivalence class, all the configurations that are obtained one from the other by a translation. We obtain that the different equivalence classes can be identified for example reading the sequence starting from particle $1$ and moving anticlockwise. An equivalence class is then identified by a permutation of $2,3,\dots ,N$ and consequently we have $(N-1)!$ equivalence classes.

\subsection{The alphabet model on the interval}
The model is much easier if the particles are moving on a linear lattice with $N$ sites instead that on a ring. In this case the un-oriented physical graph where the particles move has $V=\{1,2,\dots ,N\}$ with edges $\{i,i+1\}$, $i=1,\dots ,N-1$. In this case given any cycle $\mathcal C$ in the graph of configurations  we have for any pair of particles $i$ and $j$ that $\mathcal N_{i,j}-\mathcal N_{j,i}=0$. Using the first two equalities in \eqref{itad} that are not related to the special geometry of the graph we obtain that in this case the Kolmogorov criterion is always satisfied.

\subsection{The k-types case}

The cases when the types of particles evolving are $k<N$ is just a special case of the alphabet case. Indeed if we consider for example the alphabet model with $\gamma(i,j)=0$ and $\gamma(i,\cdot)=\gamma(j,\cdot)$ the model that we obtain is a model with $N-1$ types of particles since the particles $i$ and $j$ become particles of the same type. Fixing in particular $c(i,j)=c(j,i)=0$ the two labeled particles cannot exchange their position and the model is exactly equivalent to a model with two unlabeled identical particles.

If we have $n_1$ particles of type $1$, $n_2$ particles of type 2 up to $n_k$ particles of type $k$ then we obtain as a condition for reversibility on the ring
\begin{equation}\label{vacaz}
\sum_{j\neq i}n_j\gamma(i,j)=0\,, \qquad \forall i=1,\dots ,k\,.
\end{equation}
A special case of interest is when on a ring there are $k$ types of particles, $N$ is divisible by $k$ and there are exactly $\frac Nk$ particles for each type. In this case the reversibility condition on the ring is again the zero divergence condition \eqref{rev} on the graph of types that has $k$ nodes.

On the interval, as in the alphabet case, the model is always reversible.

\section{The energy}\label{ensec}
We discuss in detail the energy associated to the invariant measure like in \eqref{mis-dav}. In particular we show that there are simple and natural representations with direct combinatorial interpretations.

\subsection{The alphabet model on the ring}
For a reversible alphabet model the energy associated to a given configuration can be computed as follows. Fix a reference configuration $\eta^*$. This can be for example the configuration with particle $i$ at site $i$. The energy of $\eta^*$ is fixed as zero. Fix any path $(\eta^{(0)},\eta^{(1)},\dots ,\eta^{(n)})$ on the configuration space with $\eta^{(0)}=\eta^*$ and $\eta^{(n)}=\eta$. Then we have
\begin{equation}\label{via}
H_N(\eta)=-\sum_{m=0}^{n-1}\log\frac{r(\eta^{(m)},\eta^{(m+1)})}{r(\eta^{(m+1)},\eta^{(m)})}\,.
\end{equation}
Indeed since $H_N(\tau_x\eta^*)=H_N(\eta^*)$ the initial configuration of the path can be any configuration obtained simply shifting all the particles of $\eta^*$. A possible choice of the starting configuration to compute the energy of a configuration $\eta$ is then obtained keeping fixed particle $1$ along the path. This corresponds to consider as a reference starting configuration $\tau_{x_{\eta}(1)-1}\eta^*$. The path can be realized forbidding exchanges on the bonds $(x_{\eta}(1),x_{\eta}(1)+1)$ and $(x_{\eta}(1)-1,x_{\eta}(1))$. In this way we are moving the remaining $N-1$ particles on an interval and in particular when $1<i<j$ we have that along any path $\mathcal N_{i,j}-\mathcal N_{j,i}$ is $0$ if $x_\eta(j)\not\in[x_\eta(1),x_\eta(i)]$ while otherwise it is equal to $-1$. We then have using the third expression in formula \eqref{itad}
\begin{equation}\label{io}
H_N(\eta)=\sum_{1<i<j\leq N}\gamma(i,j)\chi_{[x_\eta(1),x_\eta(i)]}(x_\eta(j))\,,
\end{equation}
where $\chi_S(\cdot)$ is the characteristic function of the set $S$.

Another possible choice is to keep fixed the reference configuration choosing in particular $\eta^*$ as the configuration on which particle labeled with $i$ is in the $i$ site . The path can be realized avoiding for example to have exchanges across the edge
$\{N,1\}$. In this way the problem is equivalent to a problem on an interval with $N$ particles. Any path that connects the reference configuration $\eta^*$ to a given configuration $\eta$ will have the following property. Consider $i<j$. If in the configuration $\eta$ the $i$ particle occupies a site with a label bigger than the one of the particle $j$ then we have that $\mathcal N_{i,j}-\mathcal N_{j,i}=-1$ while instead if the $i$ particle occupies a site with a label smaller than the one of the particle $j$ then we have that $\mathcal N_{i,j}-\mathcal N_{j,i}=0$. Then still using the third expression of \eqref{itad} we obtain an expression equivalent to \eqref{io} that is
\begin{equation}\label{copa}
H_N(\eta)=\sum_{1\leq i<j\leq N}\gamma(i,j)\chi_{[1,x_\eta(i)]}(x_\eta(j))\,.
\end{equation}
This is a generalization of \eqref{prima-ham-dav}.
\begin{remark}\label{remarco}
Formulas \eqref{io} and \eqref{copa} hold also for the k types case. If for example particles $i$ and $j$ are of the same type it is enough to consider $\gamma(i,j)=0$, $\gamma(i,\cdot)=\gamma(j,\cdot)$. It is not difficult to show that if we have $n_1$ particles of type $1$, $n_2$ particles of type 2 up to $n_k$ particles of type $k$ and the reversibility condition \eqref{vacaz} is satisfied
then the energy \eqref{copa} can be written as a sum on the $N$ sites of the ring as
\begin{equation}\label{esia}
\sum_{1\leq x<y\leq N}\gamma\left(\tilde\eta(y),\tilde\eta(x)\right)\chi_{(0,+\infty)}\left(\tilde\eta(x)-\tilde\eta(y)\right)\,.
\end{equation}
We are not going to discuss in detail this simple computation. We are going instead to discuss in the following sections the special cases when there are $\frac{N}{k}$ particles of each type. In these cases we obtain formulas with a different structure.
\end{remark}

\begin{remark}
Formula \eqref{copa} gives also an expression for the energy of any alphabet model on an interval and likewise formula \eqref{esia} gives also the energy for any $k$ types model on an interval.
\end{remark}

\smallskip

Let us analyze directly the cases with a few types of particles, let us say $3$ and $4$.
In the case of 3 particles there are just 2 equivalence classes of configurations that are
ABC and ACB. Recall that we individuate an equivalence class starting from the particle $A$ and following the word in the anticlockwise direction. We can reach a configuration in the equivalence class $ACB$ with a single transposition $CB \to BC$ starting from the reference configuration $ABC$. We set $H(ABC)=0$ and consequently $H(ACB)=\gamma(B,C)$.
In this case the dimension of $\Lambda^1_d$ is $1$, the vector of the basis is associated to the unique cycle in Figure 2 so that there is a one parameter family of reversible models parameterized by $\alpha$ and $\gamma(A,B)=\gamma(B,C)=\gamma(C,A)=\alpha$ so that $H(ACB)=\alpha$. Fixed the parameter $\alpha$ there is still some freedom in the choice of the rates and as a special case we can recover the original model described in section \ref{mod}.
\begin{figure}
\begin{tikzpicture}
\node [shape=circle,fill=black!20,draw] (A) at    (-4,0) {A};
\node [shape=circle,fill=black!20,draw] (B) at    (0,0) {B};
\node [shape=circle,fill=black!20,draw] (C) at    (-2,4) {C};

\node [shape=circle,fill=black!0] (T) at    (-2,1.4) {$\alpha$};
 \draw[->] (-1.3,1.4) arc (0:+360:0.7cm);

\draw[thick,-,color=black!80] (A) to (B) ;
\draw[thick,-,color=black!80] (B) to (C) ;
\draw[thick,-,color=black!80] (C) to (A) ;

\node [shape=circle,fill=black!20,draw] (A) at    (2,0) {1};
\node [shape=circle,fill=black!20,draw] (B) at    (6,0) {2};
\node [shape=circle,fill=black!20,draw] (C) at    (4,4) {3};
\node [shape=circle,fill=black!20,draw] (D) at    (4,1.5) {4};

\node [shape=circle,fill=black!0] (T) at    (4,0.6) {$\alpha$};
 \draw[->] (4.3,0.6) arc (0:+360:0.3cm);
\node [shape=circle,fill=black!0] (T) at    (4.5,2) {$\beta$};
 \draw[->] (4.8,2) arc (0:+360:0.3cm);
\node [shape=circle,fill=black!0] (T) at    (3.5,2) {$\delta$};
 \draw[->] (3.8,2) arc (0:+360:0.3cm);

\draw[thick,-,color=black!80] (A) to (B) ;
\draw[thick,-,color=black!80] (C) to (B) ;
\draw[thick,-,color=black!80] (A) to (C) ;
\draw[thick,-,color=black!80] (D) to (A) ;
\draw[thick,-,color=black!80] (D) to (B) ;
\draw[thick,-,color=black!80] (D) to (C) ;

\end{tikzpicture}
\caption{The graph of types for $k=3$  (left) and for $k=4$ (right) with elementary cycles of a basis of $\Lambda^1_d$.}
\label{fig:cycles}
\end{figure}
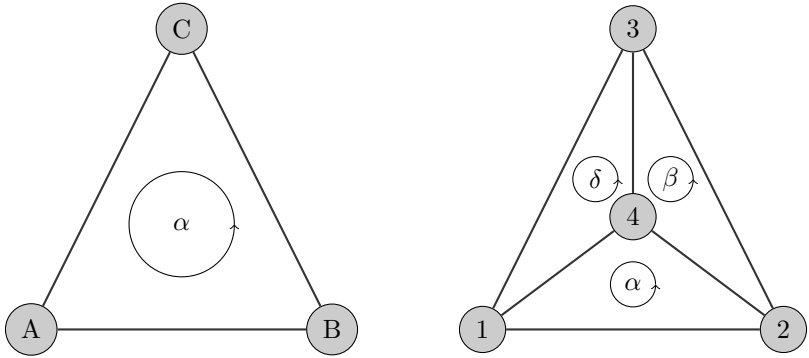

In the case of 4 particles there are $3!=6$ equivalence classes that are $1234$, $1243$, $1324$, $1342$, $1423$ and $1432$. We fix $H(1234)=0$. Then we have $H(1243)=\gamma(4,3)$, $H(1324)=\gamma(3,2)$, $H(1342)=\gamma(1,2)$ and $H(1423)=\gamma(4,1)$. The only configuration that cannot be reached from $1234$ with one single transposition is the reversed one that is $1432$ and we have for example $H(1432)=\gamma(3,4)+\gamma(1,2)$. The dimension of $\Lambda^1_d$ in this case is $3$ and 3 vectors of a basis are the ones associated to the cycles in Figure 2. If we call $\alpha, \beta$ and $\delta$ the corresponding parameters we have $H(1243)=\delta-\beta$, $H(1324)=-\beta$, $H(1342)=\alpha$ and $H(1423)=\alpha-\delta$ and $H(1432)=\beta+\alpha-\delta$.

\subsection{The ABC model}\label{abcase}

We can use these results to give for example another representation for the Hamiltonian of the ABC model on a ring with $\frac N3$ particles of each type. The advantage of this new representation is on the fact that it has a simple and natural topological interpretation, it is clearly well defined and translational invariant and it is useful for computations. Instead of writing the Hamiltonian in terms of two body long range interactions, like in \eqref{prima-ham-dav} and \eqref{ham2-dav}, we write it in terms of a
3 body long range interaction. More precisely we consider an Hamiltonian of the form
\begin{equation}\label{3body}
H_N(\eta)=\sum_{x,y,z}J(x,y,z)\eta_A(x)\eta_B(y)\eta_C(z)\,,
\end{equation}
where $J$ is defined as follows.
Given $(x,y,z)$ an ordered triple of distinct points on $\mathbb Z_N$ we say that they are \emph{well oriented} if moving on the ring anticlockwise starting from $x$ we meet $y$ before $z$ (or equivalently $y\in [x,z]$). We say instead that $(x,y,z)$ are \emph{badly oriented} if  moving on the ring anticlockwise starting from $x$ we meet $z$ before $y$. We call $\mathcal B$ the collection of badly oriented triples. The interaction appearing in \eqref{3body} has to be chosen as
\begin{equation}\label{defJ}
J(x,y,z):=\frac{3\log q}{N}\chi_\mathcal B(x,y,z)\,.
\end{equation}
Alternatively if we define
\begin{equation}\label{Bdav}
\mathcal B(\eta):=\big\{(x,y,z)\in \mathcal B\,:\, \eta_A(x)=1, \eta_B(y)=1, \eta_C(z)=1\big\}
\end{equation}
the collection of badly oriented triples of particles A, B and C,
we have that
\begin{equation}\label{22dav}
H_N(\eta)=\frac{3\log q}{N}\Big|\mathcal B(\eta)\Big|\,.
\end{equation}
To prove that \eqref{3body} is correct  we need to check the validity of detailed balance and for this we have to compute $H_N(\eta^{x,x+1})-H_N(\eta)$.
Consider for example the case $\eta_A(x)=1$ and $\eta_B(x+1)=1$, the other cases can be discussed similarly.
The exchanged configuration will have
$\eta^{x,x+1}_B(x)=1$ and $\eta^{x,x+1}_A(x+1)=1$. Consider any triple of particles A, B and C in the configuration $\eta$ with particles belonging to sites different from $x$ and $x+1$. The triples of this type will appear also in the configuration $\eta^{x,x+1}$ and moreover they will have the same orientation, either good or bad. This means that there will be no contribution to the difference of energy from triples of this type. Consider now a triple of particles A, B and C in the configuration $\eta$ having the particle of type A in $x$ and the particle of type B not in ${x+1}$. After the transposition only the particle A is moving of one unit in the lattice but the orientation of the triple is preserved. The same happens if we consider a triple of particles A, B and C in the configuration $\eta$ having the particle of type B in $x+1$ and the particle of type A not in ${x}$. Also triples of this type will not contribute to the difference of energy. Finally we have to consider the triples having the particle of type A in $x$ and the particle of type $B$ in $x+1$. There are exactly $\frac N3$ triples of this type, one for each C particle, and all of them are well oriented. After the transposition all of them become badly oriented so that the contribution to the increment of energy coming from configurations of this type is equal to $\log q$ and this implies the validity of detailed balance.

\begin{figure}
\begin{tikzpicture}[scale=0.8]
\draw (0,0) circle (2cm);
\node [shape=circle,fill=black!20,draw] (C) at
    ($ ({cos(90)*2},{sin(90)*2}) $) {C};
\node [shape=circle,fill=black!20,draw] (A) at
    ($ ({cos(225)*2},{sin(225)*2}) $) {A};
\node [shape=circle,fill=black!20,draw] (B) at
    ($ ({cos(0)*2},{sin(0)*2}) $) {B};

\draw[rounded corners,thick,->]
	(A)--($ ({cos(225)*2+0.5*(cos(90))},{sin(225)*2-0.5*(sin(90))}) $) arc (225:360:2.3cm)--(B);
\draw[rounded corners,thick,->]
	(B)--($ ({cos(0)*2+0.5*(cos(45))},{sin(0)*2+0.5*(sin(45))}) $) arc (0:90:2.3cm)--(C);
	\draw[rounded corners,thick,->]
	(C)--($ ({cos(90)*2+0.5*(cos(135))},{sin(90)*2+0.5*(sin(135))}) $) arc (90:225:2.3cm)--(A);
\draw (7,0) circle (2cm);
\node [shape=circle,fill=black!20,draw] (B) at
    ($ ({7+cos(90)*2},{sin(90)*2}) $) {B};
\node [shape=circle,fill=black!20,draw] (A) at
    ($ ({7+cos(225)*2},{sin(225)*2}) $) {A};
\node [shape=circle,fill=black!20,draw] (C) at
    ($ ({7+cos(0)*2},{sin(0)*2}) $) {C};
\draw[rounded corners,thick,->]
	(A)--($ ({7+cos(225)*2+0.5*(cos(90))},{sin(225)*2-0.5*(sin(90))}) $) arc (225:450:2.6cm)--(B);
\draw[rounded corners,thick,->]
	(B)--($ ({7+cos(90)*2+0.5*(cos(90))},{sin(90)*2-0.5*(sin(90))}) $) arc (90:360:1.3cm)--(C);
\draw[rounded corners,thick,->]
	(C)--($ ({7+cos(0)*2+0.5*(cos(45))},{sin(0)*2+0.5*(sin(45))}) $) arc (0:225:2.6cm)--(A);
\end{tikzpicture}
\caption{Going from A to B to C and then again to A anticlockwise we get $N$ if the triple is well oriented (left)
and $2N$ if it is badly oriented (right). }
\label{fig:connections}
\end{figure}
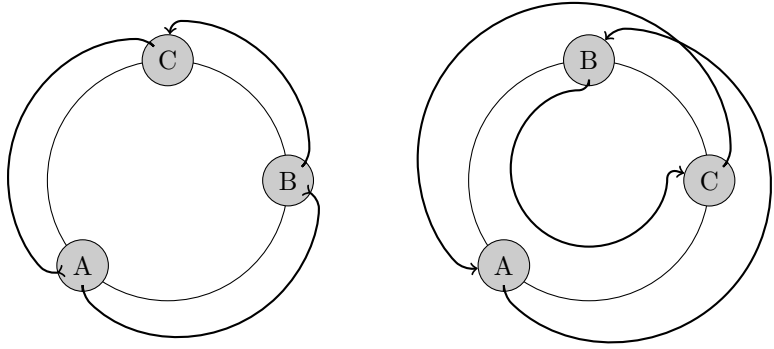

Another nice way of proving that \eqref{3body} is correct is by showing that up to an additive constant it coincides with \eqref{ham2-dav}. This follows directly observing that for any triple of particles A, B, C summing the length obtained going anticlockwise from A to B to C and then back to A we obtain $N$ if the triple is well ordered and $2N$ if the triple is badly ordered (see Figure \ref{fig:connections})

\subsubsection{Large deviations}\label{ldpabc}
A natural and interesting regime of the system is the weakly asymmetric one \cite{CDE} in which the number of sites $N$ is diverging
and the asymmetry parameter $q$ is chosen to depend on $N$ in the form $q_N= e^{-\frac \beta N}$ where $\beta$ is a
positive parameter. We restrict to the case of equal number of particles. Let us define the empirical measure $\pi_N(\eta)=\left(\pi_N^A(\eta),\pi_N^B(\eta),\pi_N^C(\eta)\right)$. This is a triple of positive measures on the unitary circle $S^1=\mathbb R/\mathbb Z$ defined by
\begin{equation}\label{emp-dav}
\pi_N^X(\eta):=\frac 1N\sum_{x\in \mathbb Z_N}\eta_X(x)\delta_{\frac xN}\,,\qquad X=A,B,C\,.
\end{equation}
In this regime, following the arguments in \cite{BCP} but using the Hamiltonian \eqref{3body}, we can prove a large deviations principle for the empirical measure
\begin{equation}\label{ldpdav}
\mathbb P_{\mu_N}\left(\pi_N(\eta)\sim \rho\right)\simeq e^{-NV(\rho)}\,.
\end{equation}
The rate functional $V(\rho)$ is
\begin{equation}\label{LDP-rate-dav}
\sum_{X=A,B,C}\int_{0}^1\rho_X(x)\log 3\rho_X(x) d\,x+3\beta\int_{\mathcal B}\rho_A(x)\rho_B(y)\rho_C(z)\, dx\, dy\, dz+\textrm{constant}\,,
\end{equation}
provided that the total mass of the three densities is equal to $\frac 13$.
We used the same notation $\mathcal B$ for the collection of badly oriented triples of points on the unitary continuous ring $S^1$ whose definition is exactly the same as in the discrete ring.
This rate functional must coincide with the one in \cite{BCP} that differs from \eqref{LDP-rate-dav} (apart a different additive constant) only in the second term that in \cite{BCP} reads
\begin{align}
& \beta\int_0^1dx\int_x^1dy\left[\rho_A(x)\rho_C(y)+\rho_B(x)\rho_A(y)+\rho_C(x)\rho_B(y)\right]\nonumber \\
& =\frac \beta3 -\beta\int_0^1dx\left[\rho_A(x)F_C(x)+\rho_B(x)F_A(x)+\rho_C(x)F_B(x)\right]\label{es-dav}
\end{align}
We are considering $\rho_A$, $\rho_B$ and $\rho_C$ as periodic functions of period one on all the real line and we call $F_X(x)=\int_0^x \rho_X(y)dy$ where $X$ can be A or B or C. This curious identity can be shown with some tricky integrations by parts. This computation is the continuous counterpart of the previous discrete ones.
Indeed the second term in \eqref{LDP-rate-dav} can be written as
\begin{align}
& 3\beta\int_0^1\rho_A(x)dx\int_x^{x+1}\rho_C(y)dy\int_y^{x+1}\rho_B(z)dz\nonumber\\
&  =3\beta\int_0^1\rho_A(x)\left\{\Big[F_C(y)\int_y^{x+1}\rho_B(z)dz\Big]_x^{x+1}+\int_x^{x+1}F_C(y)\rho_B(y)dy\right\}\nonumber\\
& =-\beta\int_0^1\rho_A(x)F_C(x)dx+3\beta\int_0^1\rho_A(x)dx\int_x^{x+1}F_C(y)\rho_B(y)dy \label{conto-dav}
\end{align}
Observe that
\begin{equation}\label{oss-dav}
\frac{d}{dx}\int_x^{x+1}F_C(y)\rho_B(y)dy=\frac 13 \rho_B(x)\,.
\end{equation}
The second term in the right hand side of \eqref{conto-dav} can be transformed as
\begin{align}
& 3\beta\int_0^1\rho_A(x)dx\int_x^{x+1}F_C(y)\rho_B(y)dy \nonumber \\
& =\beta\Big[3F_A(x)\int_x^{x+1}F_C(y)\rho_B(y)dy\Big]_0^1-\beta\int_0^1\rho_B(x)F_A(x)dx \nonumber \\
& =\beta\int_1^2F_C(x)\rho_B(x)dx-\beta\int_0^1\rho_B(x)F_A(x)dx\nonumber\\
& =\frac{2\beta}{9} -\beta\int_0^1\rho_C(x)F_B(x)dx-\int_0^1\rho_B(x)F_A(x)dx
\end{align}
Putting all together we obtain that \eqref{LDP-rate-dav} is equal to the one written in \cite{BCP}.

\subsection{The k-types case}\label{ktypesec}
Formula \eqref{3body} can be generalized to the case of k types of particles. The representation that we obtain has the advantage that is very useful for  computations with respect to the usual representation in terms of 2 body interactions. We consider k types of particles evolving on a ring with $N$ sites and such that there are exactly $\frac{N}{k}$ particles for each type. In this case as we already discussed the reversibility of the model coincides with the condition of zero divergence for the discrete vector field $\gamma$
on the graph of types with $k$ vertices. The energy for this system of particles can be written naturally in terms of a $k$ body long range interaction
\begin{equation}\label{kbody}
H_N(\eta)=\sum_{x_1,\dots,x_k}J(x_1,\dots,x_k)\eta_1(x_1)\dots\eta_k(x_k)\,.
\end{equation}
The sum in \eqref{kbody} is over distinct ordered collections $\underline x:=(x_1,\dots ,x_k)$
of $k$ elements of the ring $\mathbb Z_N$.
The interaction $J$ is constructed as follows. Since $\gamma$ is divergence free on the graph of types with $k$ vertices we can consider a reversible alphabet model on a ring with $k$ sites and having the same $\gamma$. We call $\xi=\left(\xi(x)\right)_{x\in \mathbb Z_k}$ a generic configuration of this alphabet model on a ring with $k$ sites.  The  energy is well defined and we call $H_k(\xi)$ the energy of the configuration $\xi$ of the alphabet model with $k$ particles. Two configurations in the same equivalence class (coinciding modulo translations) have the same energy. Consider $\underline x:=(x_1,\dots,x_k)$ an ordered collection of $k$ distinct point of the ring with $N$ sites. We put a particle of type $1$ in $x_1$, a particle of type $2$ on $x_2$ up to a particle of type $k$ on $x_k$.
We associate to $\underline x$ the configuration $\xi^{\underline x}$ of the $k$ alphabet model defined as follows. Let $\pi$ be a permutation on $\{1,2,\dots ,k\}$ such that moving anticlockwise on the ring $\mathbb Z_N$ starting from $x_{\pi(1)}$ we meet before $x_{\pi(2)}$
then $x_{\pi(3)}$ up to $x_{\pi(k)}$ exactly in this order before coming back to $x_{\pi(1)}$. We define $\xi^{\underline x}$ by the condition $\tilde\xi^{\underline x}=\left(\pi(1),\pi(2),\dots ,\pi(k)\right)$. Recall that given a configuration $\xi$ we have that $\tilde\xi(x)=j$ if at site $x$ there is a particle of type $j$. Since the permutation $\pi$ is not univocally determined then also the configuration $\xi^{\underline x}$ is not unique but all of them are obtained one from the other by a global shift so that all of them belong to the same equivalence class. In particular $H_k\left(\xi^{\underline x}\right)$ is univocally determined. Essentially what we are doing is to observe just the relative positions of the $k$ particles in $\underline x$ neglecting the remaining sites.  We have
\begin{equation}\label{conk}
J(x_1,\dots,x_k)=\left(\frac{k}{N}\right)^{k-2}H_k\left(\xi^{\underline x}\right)\,.
\end{equation}
We can also write a formula similar to \eqref{22dav}. Given $\eta$ a configuration of particles in $\mathbb Z_N$ and $\left[\xi\right]$ an equivalence class of configurations modulo translations for the $k$ alphabet process, we call $\mathcal N_{\left[\xi\right]}(\eta)$ the number of $k-$uples of particles of different types in $\eta$ that are relatively arranged on the ring like in the equivalence class $\left[\xi\right]$. The energy \eqref{kbody} can then be written like
\begin{equation}\label{kaltra}
H_N(\eta)=\left(\frac kN\right)^{k-2}\sum_{\left[\xi\right]}H_k(\xi)\mathcal N_{\left[\xi\right]}(\eta)\,.
\end{equation}
The proofs of these results follow by the same arguments in section \ref{abcase}. Similar continuous formulas and identities like in section \ref{ldpabc} can be obtained also in this case. These formulas in terms of k-body interactions have a much simpler structure with respect to the 2-body interaction formulas in \cite{EKKM2, FF1, FF2}.

\section{The graph of the interaction}

We show that much of the information about the interaction between particles can be encoded in a Tournament associated to the dynamics.
We consider here the alphabet model. The tournament is on the N vertices corresponding to particle types  and is constructed as follows. Given a pair of particles $i$ and $j$ a tournament has only one between the two possible edges $(i,j)$ and $(j,i)$. We associate the orientation corresponding to the negative value of $\gamma$. This means that we take $(i,j)$ if $c(i,j)<c(j,i)$ and we take $(j,i)$ otherwise. We are not going to consider the cases when $\gamma(i,j)=0$ for some $i,j$. With $N$ vertices it is possible to construct $2^{\frac{N(N-1)}{2}}$ different tournaments but not all of them can be obtained in this way from a reversible alphabet model.

We claim that the tournament associated to a reversible alphabet model is necessarily strong. This does not mean that if a model has a tournament that is strong then it is reversible. However it is always possible to define some rates in such a way that we have a reversible model with associated a fixed strong tournament. This follows from Lemma \ref{IHPqui}.

Consider indeed a reversible alphabet model and let $(V,E)$ be the corresponding tournament. We have that $(V,\mathcal E)$ is the complete graph and is therefore connected. For any $(i,j)\in E$ consider the flow $Q(i,j):=-\gamma(i,j)$. This is a non zero flow on $(V,E)$ such that $\div Q=-\div \gamma$. By reversibility \eqref{rev}
we have then that $Q$ is divergence free. By Lemma \ref{IHPqui} the tournament $(V,E)$ is strong.

Conversely consider a strong tournament $(V,E)$. Then by Lemma \ref{IHPqui} there exists a positive divergence free flow $Q$. Let us define $\gamma(i,j):=-Q(i,j)$ if $(i,j)\in E$ and $\gamma(i,j)=Q(j,i)$ if $(i,j)\not\in E$. Fixing the rates according to \eqref{gamma} we obtain a reversible model with associated the tournament $(V,E)$.

\subsection{Examples of models with not strong tournaments}

In the following we will discuss models associated to strong tournaments. We start however giving examples of models associated
to tournaments that are not strong.
The first example is a collection of solvable models. Consider $F_1, \dots ,F_N$ some real numbers. We interpret $F_i$ as the external field that act on particle $i$. We assume that all of them are different and that $|F_i|+|F_j|<1$ for any pair $i,j$. We define the rates of exchange as
\begin{equation}\label{concampo}
c(i,j)=1+F_i-F_j\,.
\end{equation}
Since all the external fields are different the assignment of an external field to each particle induces a total order among the particles obtained comparing the corresponding values. The tournament $(V,E)$ associated to the rates \eqref{concampo} is such that $(i,j)\in E$ when $F_i<F_j$. Such a tournament cannot be strong. The vertex associated to the maximal value of the field, for example, has all the edges oriented toward it. The model is then necessarily not reversible. Let us now show that for any choice of the external fields the uniform measure over all possible configurations of particles is invariant. This corresponds to check the validity of
\begin{equation}\label{stF}
\sum_{x=1}^Nr(\eta,\eta^{x,x+1})=\sum_{x=1}^Nr(\eta^{x,x+1},\eta)\,, \qquad \forall \eta\,.
\end{equation}
Using definition \eqref{concampo} equation \eqref{stF} becomes
\begin{equation}\label{stFs}
N+\sum_{x=1}^N\left(F_{\tilde\eta(x)}-F_{\tilde\eta(x+1)}\right)=N+\sum_{x=1}^N\left(F_{\tilde\eta(x+1)}-F_{\tilde\eta(x)}\right)\,,
\end{equation}
where $\tilde \eta$ is defined by setting $\tilde\eta(x)=i$ if at $x$ there is the particle $i$. Equation \eqref{stFs} is satisfied since the two sums on the two sides are telescoping and are identically zero.

Clearly the combinatorial structure
of the tournament is not identifying completely the model and we can have different models associated to the same tournament, with different invariant measures.
For example we can consider also the $N$-priority ASEP \cite{SB} or the totally asymmetric exclusion process with particles of different classes \cite{A,FM}. The N priority ASEP both on a ring or on an interval with N sites is defined by \eqref{gendav}, \eqref{exrra} with
\begin{equation}\label{ggnp}
c(i,j):=\left\{
\begin{array}{ll}
q^{-1}  & \textrm{if}\ i<j\,, \\
q & \textrm{if}\ i>j\,,
\end{array}
\right.
\end{equation}
where $q$ is a fixed positive parameter that we assume smaller than one.
The TASEP with N classes of particles both on a ring or on an interval with N sites is defined by \eqref{gendav}, \eqref{exrra} with
\begin{equation}\label{ggn}
c(i,j):=\left\{
\begin{array}{ll}
1  & \textrm{if}\ i<j\,, \\
0 & \textrm{if}\ i>j\,.
\end{array}
\right.
\end{equation}
The tournaments associated to these models are never strong and are similar to the ones of the exactly solvable models discussed above. In particular given two labels of the particles $i,j$ with $i<j$ then the oriented edge of the tournament will be $(j,i)$. In particular all the arrows are oriented toward the vertex $1$ and the tournament can not be strong.
On the ring the two models are then not reversible. For the N priority ASEP the invariant measure is not known while for the TASEP with
N classes of particles the invariant measure has an interesting combinatorial structure \cite{A,FM}.

On the interval with $N$ sites the energy of the N-priority TASEP can be written like \eqref{copa} with $\gamma(i,j)=-2\log q$ when $i<j$. The multiclass TASEP is instead not irreducible and has an unique attracting state that corresponds to have
all the particles arranged in decreasing order with respect to the labels.
By Remark \ref{remarco} the energy of a k priority ASEP on an interval of $N$ sites is given by \eqref{esia} with the $\gamma$ as before. This is exactly the statement $(i)$ of Theorem 3.1 in \cite{SB}.

\subsection{A probabilistic proof of Camion Theorem}\label{camionsec}
Using the results obtained so far we give a probabilistic proof of Camion Theorem that states that any strong tournament is Hamiltonian. The converse statement is also true but its proof is trivial.
The proof is as follows.
Consider a strong tournament and construct as above an associated reversible alphabet model. Let $H_N$ be the corresponding energy associated to the invariant measure. Since the state space of the alphabet model is finite there exists at least one configuration of particles $\eta$ that is a minimizer of $H_N$. This minimizer has to satisfies the condition
\begin{equation}\label{condag}
\gamma(\tilde\eta(x),\tilde\eta(x+1))<0\,, \qquad \forall x\,.
\end{equation}
This is because if this is not the case then we have
$$H_N(\eta^{x,x+1})-H_N(\eta)=-\gamma(\tilde\eta(x),\tilde\eta(x+1))<0$$
and this is impossible since $\eta$ is a minimizer of the energy. Condition \eqref{condag} implies that the cycle $(\tilde\eta(1),\tilde\eta(2),\dots ,\tilde\eta(N),\tilde\eta(1))$ is an Hamiltonian cycle in the tournament. Recall that $\tilde\eta(x)=i$ when at $x$ there is a particle of type $i$.

\smallskip
Using the same type of arguments it is possible to obtain also a proof of the fact that every tournament is traceable. In this case we need to construct an alphabet model on the interval.
Consider a tournament and define values of $\gamma$ compatible with the orientation. Construct an alphabet model on the interval associated to this $\gamma$. Since any alphabet model on the interval is reversible it will have an invariant measure like \eqref{mis-dav} with associated an energy like \eqref{copa}. Since the state space if finite this energy has to have at least one minimizer. This minimizer has to satisfies condition \eqref{condag} for $x=1,\dots, N-1$ and this is exactly equivalent to require that $(\tilde\eta(1),\tilde\eta(2),\dots ,\tilde\eta(N))$ is an Hamiltonian path in the  tournament.

\section{Minimizers}
We discuss the minimizers of the alphabet model with small $N$ and use this analysis to understand the structure of the minimizers
for the k types case with small k and $\frac{N}{k}$ particles of each type.

\subsection{The alphabet model}
The energy of the alphabet model has a nice combinatorial structure and it is an interesting problem to study the minimizers.
As already discussed much of the information about the interactions among the particles is encoded in the associated tournament. In particular the maximal number of minimizers and which configurations can be possible minimizers can be deduced directly from the structure of the tournament. Which of these possible minimizers are indeed true minimizers depends instead from the numerical values of the coefficients $\gamma$. The basic ingredients that we need have been discussed in the previous sections and are the following.
The first fact is that only strong tournaments arise as possible tournaments for reversible alphabet models and a quite detailed classification of tournaments (see \cite{Moon}) can be done explicitly in the case of a small number of type of particles. The second fact is that a configuration $\eta$ is a possible minimizer only if $(\tilde\eta(1),\tilde\eta(2),\dots ,\tilde\eta(N),\tilde\eta(1))$ is an Hamiltonian cycle in the corresponding tournament. Indeed if this is not the case and for example $(\tilde\eta(y),\tilde\eta(y+1))$ is not an edge of the tournament in the graph of types then $\eta^{y,y+1}$ satisfies $H_N(\eta^{y,y+1})<H_N(\eta)$.

\subsection{The k types case}
 What discussed above is relevant also for systems with a large number of particles. In particular for the cases with $k=3,4,5$ types of particles on a ring with $N$ sites and $\frac Nk$ particles of each type. This is because the energy for such a system is strongly related to the energy of the k alphabet model with the same $\gamma$ as shown in sections \ref{abcase} and \ref{ktypesec}.

In this case we show first of all that if we are interested just in the minimum value of the energy we can restrict to configurations having a block structure. This means configurations where all the particles of each single type  are grouped together in single clusters.
In particular we claim that there exists always a minimizer of the energy of this type. However for some specific models we will show that there are also minimizers not of this type.

Our claim follows by formulas \eqref{kbody}, \eqref{conk} and \eqref{kaltra}. They say that the energy is always obtained as a sum of
$\left(\frac Nk\right)^k$ terms. This is exactly the number of ways in which it is possible to select one particle of type $1$, one particle of type $2$ and so on up to one particle of type $k$. Each of these terms that we are summing can be equal to $H_k\left(\left[\xi\right]\right)$ where $\left[\xi\right]$ is a possible equivalence class of the $k$ alphabet model. Let $\left[\xi^*\right]$ a minimizer of $H_k$. We construct a configuration $\eta^*$ on $\mathbb Z_N$ simply substituting each particle in $\left[\xi^*\right]$ with a cluster of $\frac Nk$ particles of the same type. Whatever is our choice of one particle for each type we obtain particles relatively organized in the ring like the equivalence class $\left[\xi^*\right]$ and all the $\left(\frac Nk\right)^k$ terms will be equal to $H_k\left(\left[\xi^*\right]\right)$ that is the minimal one. This implies that $\eta^*$ is necessarily a minimizer.

\subsection{The $k=2$ case}
\begin{figure}
\begin{tikzpicture}
\node [shape=circle,fill=black!20,draw] (A) at    (0,-3) {};
\node [shape=circle,fill=black!20,draw] (B) at    (0,0) {};
\draw[thick,->,color=black!80] (B) to (A) ;
\end{tikzpicture}
\caption{The unique unlabeled tournament with two vertices}
\label{fig:tour2Ver}
\end{figure}
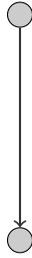

This is an elementary case and there is only one possible type of dynamics that coincides with an asymmetric exclusion. If $\gamma\neq 0$ the model on the ring cannot be reversible. This follows directly by the fact that there is only one unlabeled tournament with two sites (Figure \ref{fig:tour2Ver}) and it is not strong. Indeed this dynamics is a special case of the one defined by the rate \eqref{concampo}. The invariant measure is indeed uniform.

\subsection{The $k=3$ case}
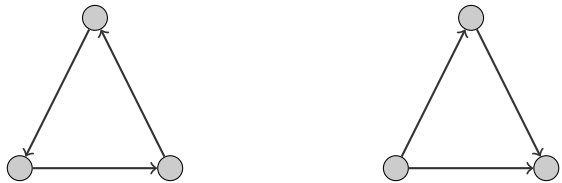
\begin{figure}
\begin{tikzpicture}
\node [shape=circle,fill=black!20,draw] (A) at    (-3,0) {};
\node [shape=circle,fill=black!20,draw] (B) at    (-1,0) {};
\node [shape=circle,fill=black!20,draw] (C) at    (-2,2) {};

\draw[thick,->,color=black!80] (A) to (B) ;
\draw[thick,->,color=black!80] (B) to (C) ;
\draw[thick,->,color=black!80] (C) to (A) ;

\node [shape=circle,fill=black!20,draw] (A) at    (2,0) {};
\node [shape=circle,fill=black!20,draw] (B) at    (4,0) {};
\node [shape=circle,fill=black!20,draw] (C) at    (3,2) {};

\draw[thick,->,color=black!80] (A) to (B) ;
\draw[thick,->,color=black!80] (C) to (B) ;
\draw[thick,->,color=black!80] (A) to (C) ;

\end{tikzpicture}
\caption{The two possible unlabeled tournament with 3 vertices. The one on the left is strong while the one on the right is not}
\label{fig:tour3Ver}
\end{figure}

In this case there are just two classes of non equivalent tournaments that are drawn in Figure \ref{fig:tour3Ver}.
Only the one on the left is strong and it is the only topological structure of interactions corresponding to a reversible alphabet model with 3 particles. The unique Hamiltonian cycle is obtained going around the triangle anticlockwise. The dimension of $\Lambda^1_d$ is 1 and there is only a one parameter family of $\gamma$. A special case is the classic ABC model.

\subsection{The $k=4$ case}
Starting from this case we draw tournaments using the convention in \cite{Moon}. In particular we are not drawing all the edges. If there is an edge missing between two vertices this means that the edge that is not drawn is oriented from the higher vertex to the lower one. In this way it is easier to understand  the topological structure of the graph looking at the picture. In the case of 4 types of particles there are 4 classes of non-isomorphic tournaments that are drawn in Figure \ref{fig:tour4Ver}.
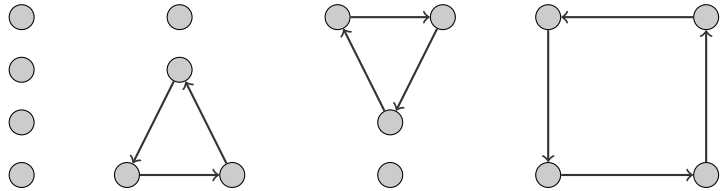
\begin{figure}
\begin{tikzpicture}[scale=0.7]
\node [shape=circle,fill=black!20,draw] (A) at    (-6,3) {};
\node [shape=circle,fill=black!20,draw] (A) at    (-6,2) {};
\node [shape=circle,fill=black!20,draw] (A) at    (-6,1) {};
\node [shape=circle,fill=black!20,draw] (A) at    (-6,0) {};


\node [shape=circle,fill=black!20,draw] (A) at    (-3,3) {};
\node [shape=circle,fill=black!20,draw] (A) at    (-4,0) {};
\node [shape=circle,fill=black!20,draw] (B) at    (-2,0) {};
\node [shape=circle,fill=black!20,draw] (C) at    (-3,2) {};

\draw[thick,->,color=black!80] (A) to (B) ;
\draw[thick,->,color=black!80] (B) to (C) ;
\draw[thick,->,color=black!80] (C) to (A) ;


\node [shape=circle,fill=black!20,draw] (A) at    (0,3) {};
\node [shape=circle,fill=black!20,draw] (B) at    (2,3) {};
\node [shape=circle,fill=black!20,draw] (C) at    (1,1) {};
\node [shape=circle,fill=black!20,draw] (D) at    (1,0) {};

\draw[thick,->,color=black!80] (A) to (B) ;
\draw[thick,->,color=black!80] (B) to (C) ;
\draw[thick,->,color=black!80] (C) to (A) ;

\node [shape=circle,fill=black!20,draw] (A) at    (4,3) {};
\node [shape=circle,fill=black!20,draw] (B) at    (7,3) {};
\node [shape=circle,fill=black!20,draw] (C) at    (7,0) {};
\node [shape=circle,fill=black!20,draw] (D) at    (4,0) {};

\draw[thick,->,color=black!80] (A) to (D) ;
\draw[thick,->,color=black!80] (D) to (C) ;
\draw[thick,->,color=black!80] (C) to (B) ;
\draw[thick,->,color=black!80] (B) to (A) ;

\end{tikzpicture}
\caption{The 4 possible unlabeled tournament with 4 vertices. Only the rightmost is strong. }
\label{fig:tour4Ver}
\end{figure}
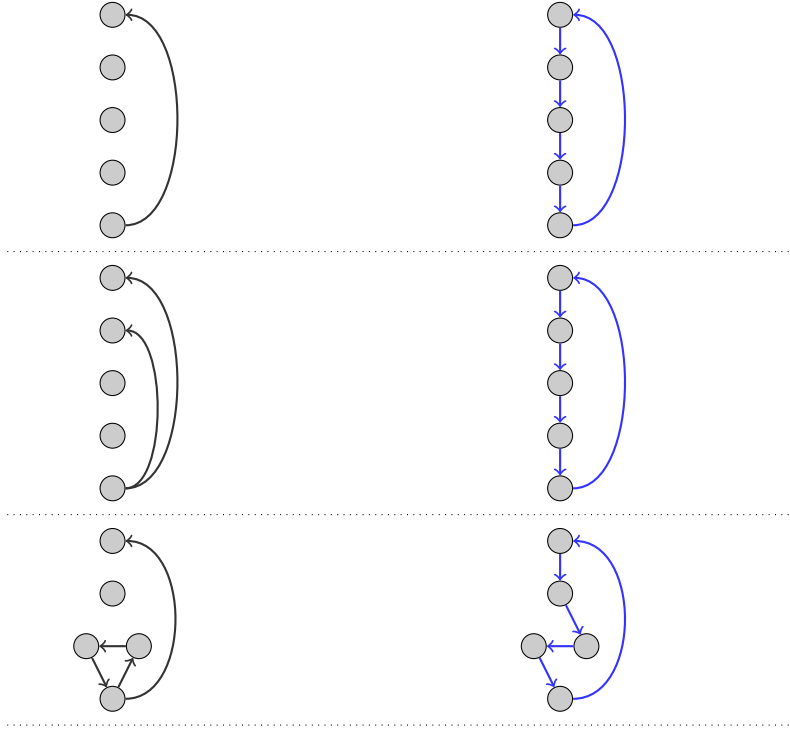
\begin{figure}
\begin{tikzpicture}[scale=0.7]

\node [shape=circle,fill=black!20,draw] (A) at      (0, 2) {};
\node [shape=circle,fill=black!20,draw] (B) at      (0, 1) {};
\node [shape=circle,fill=black!20,draw] (C) at      (0, 0) {};
\node [shape=circle,fill=black!20,draw] (D) at      (0,-1) {};
\node [shape=circle,fill=black!20,draw] (E) at      (0,-2) {};

\draw[thick,->,color=black!80] (E).. controls (1.555,-2) and (1.555,2 ) .. (A) ;

\node [shape=circle,fill=black!20,draw] (A) at     ($ ({8.5},{2}) $) {};
\node [shape=circle,fill=black!20,draw] (B) at    ($ ({8.5},{1}) $) {};
\node [shape=circle,fill=black!20,draw] (C) at    ($ ({8.5},{0}) $) {};
\node [shape=circle,fill=black!20,draw] (D) at    ($ ({8.5},{-1}) $) {};
\node [shape=circle,fill=black!20,draw] (E) at    ($ ({8.5},{-2}) $) {};
\draw[thick,->,color=blue!80] (A) to (B) ;
\draw[thick,->,color=blue!80] (B) to (C) ;
\draw[thick,->,color=blue!80] (C) to (D) ;
\draw[thick,->,color=blue!80] (D) to (E) ;
\draw[thick,->,color=blue!80] (E) .. controls (10.055,-2) and (10.055,2 ) .. (A) ;

\draw[dotted,color=black] (-2,-2.5) to (13,-2.5);

\node [shape=circle,fill=black!20,draw] (A) at      (0, -3) {};
\node [shape=circle,fill=black!20,draw] (B) at      (0, -4) {};
\node [shape=circle,fill=black!20,draw] (C) at      (0, -5) {};
\node [shape=circle,fill=black!20,draw] (D) at      (0,-6) {};
\node [shape=circle,fill=black!20,draw] (E) at      (0,-7) {};

\draw[thick,->,color=black!80] (E).. controls (1.555,-7) and (1.555,-3 ) .. (A) ;
\draw[thick,->,color=black!80] (E).. controls (1.055,-7) and (1.055,-4 ) .. (B) ;

\node [shape=circle,fill=black!20,draw] (A) at     ($ ({8.5},{-3}) $) {};
\node [shape=circle,fill=black!20,draw] (B) at    ($ ({8.5},{-4}) $) {};
\node [shape=circle,fill=black!20,draw] (C) at    ($ ({8.5},{-5}) $) {};
\node [shape=circle,fill=black!20,draw] (D) at    ($ ({8.5},{-6}) $) {};
\node [shape=circle,fill=black!20,draw] (E) at    ($ ({8.5},{-7}) $) {};
\draw[thick,->,color=blue!80] (A) to (B) ;
\draw[thick,->,color=blue!80] (B) to (C) ;
\draw[thick,->,color=blue!80] (C) to (D) ;
\draw[thick,->,color=blue!80] (D) to (E) ;
\draw[thick,->,color=blue!80] (E) .. controls (10.055,-7) and (10.055,-3 ) .. (A) ;

\draw[dotted,color=black] (-2,-7.5) to (13,-7.5);

\node [shape=circle,fill=black!20,draw] (A) at      (0, -8) {};
\node [shape=circle,fill=black!20,draw] (B) at      (0, -9) {};
\node [shape=circle,fill=black!20,draw] (C) at      (-0.5, -10) {};
\node [shape=circle,fill=black!20,draw] (D) at      (0.5,-10) {};
\node [shape=circle,fill=black!20,draw] (E) at      (0,-11) {};

\draw[thick,->,color=black!80] (D) to (C) ;
\draw[thick,->,color=black!80] (C) to (E) ;
\draw[thick,->,color=black!80] (E) to (D) ;
\draw[thick,->,color=black!80] (E).. controls (1.5055,-11) and (1.5055,-8 ) .. (A) ;

\node [shape=circle,fill=black!20,draw] (A) at      (8.5, -8) {};
\node [shape=circle,fill=black!20,draw] (B) at      (8.5, -9) {};
\node [shape=circle,fill=black!20,draw] (C) at      (8, -10) {};
\node [shape=circle,fill=black!20,draw] (D) at      (9,-10) {};
\node [shape=circle,fill=black!20,draw] (E) at      (8.5,-11) {};

\draw[thick,->,color=blue!80] (A) to (B) ;
\draw[thick,->,color=blue!80] (B) to (D) ;
\draw[thick,->,color=blue!80] (D) to (C) ;
\draw[thick,->,color=blue!80] (C) to (E) ;
\draw[thick,->,color=blue!80] (E).. controls (10.055,-11) and (10.055,-8 ) .. (A) ;

\draw[dotted,color=black] (-2,-11.5) to (13,-11.5);

\end{tikzpicture}
\caption{The 3 equivalence classes of isomorphic strong tournaments with 5 vertices (left, black)
having a unique Hamiltonian cycle (right, blue). }
\label{fig:tour5Ver}
\end{figure}
Only the rightmost is strong and corresponds to the interactions of reversible alphabet
models with 4 types of particles. It has an unique Hamiltonian cycle that is clearly obtained going around the square anticlockwise.
The dimension of $\Lambda^1_d$ is 3 and the values of  $\gamma$ can be fixed using $3$ degrees of freedom as discussed in section \ref{ensec}. Since there is only one strong tournament this means that for any choice of the free 3 parameters the structure of the tournament associated to the model will be always the same. What is changing depending on the values of $\gamma$ is the assignment of the labels $1,2,3,4$ to the vertices of the rightmost tournament in Figure \ref{fig:tour4Ver}.
Like in the case $k=3$ also in this case the only possible strong tournament  has a unique Hamiltonian cycle. By the arguments in section \ref{camionsec} this means that if we consider the associated k alphabet model it will have a unique (modulo translations) ground state of the energy in correspondence of the unique Hamiltonian cycle.

This uniqueness of the equivalence class  minimizing the energy is true also in $\mathbb Z_N$ with $\frac Nk$ particles of each type. The unique ground state has particles of each type grouped together in single clusters and the positions of the clusters on the ring have to be organized in such a way that their relative positions are the same of $\left[\xi^*\right]$ the unique minimizer of the corresponding $k$ alphabet model. In this configuration indeed all the  $\left(\frac Nk\right)^{k}$ terms obtained selecting one particle for each type will coincide with the minimal value $H_k\left(\left[\xi^*\right]\right)$ so that we have clearly a minimizer. Moreover any other configuration of particles will have at least one term corresponding to an equivalence class $\left[\xi\right]\neq\left[\xi^*\right]$. This term will contribute with $H_k\left(\left[\xi\right]\right)>H_k\left(\left[\xi^*\right]\right)$ and we cannot have a minimizer.

\subsection{The $k=5$ case}

In the case of 5 types of particles there are 12 non-isomorphic tournaments 6 of which are strong. The dimension of $\Lambda^1_d$ is 6 that is the number of free parameters that we can use to define a reversible alphabet model. In this case several topological structures of the interactions are possible and several configurations of the minimizers are possible. We draw on the left part of  Figure 7 the 3 possible unlabeled strong tournaments that have unique Hamiltonian cycles that are drawn in blue on the right. When the interaction among the particles is associated to a tournament of this type the situation is exactly as in the cases with 3 or 4 types of particles. The minimizer is unique up to translations. In the case of $\frac N5$ particles of each type the minimizer has the particles of each type grouped together in a single cluster. In the remaining 3 cases since there are more than one Hamiltonian cycle there can be more than one minimizer and in general which Hamiltonian cycle is related to the minimizer of the energy depends on the specific values of $\gamma$.

The first case of a strong unlabeled tournament with more than one Hamiltonian path is drawn in Figure 8. The labels to the vertices are given arbitrarily. The two Hamiltonian cycles are drawn in blue on the right.
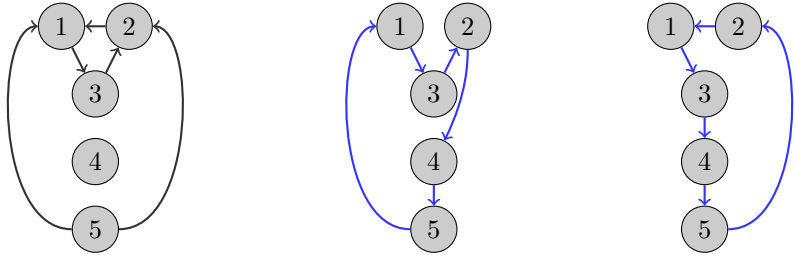
\begin{figure}
\begin{tikzpicture}[scale=0.9]
\node [shape=circle,fill=black!20,draw] (A) at      (0, 3) {1};
\node [shape=circle,fill=black!20,draw] (B) at      (1, 3) {2};
\node [shape=circle,fill=black!20,draw] (C) at      (0.5, 2) {3};
\node [shape=circle,fill=black!20,draw] (D) at      (0.5,1) {4};
\node [shape=circle,fill=black!20,draw] (E) at      (0.5,0) {5};

\draw[thick,->,color=black!80] (B) to (A) ;
\draw[thick,->,color=black!80] (A) to (C) ;
\draw[thick,->,color=black!80] (C) to (B) ;
\draw[thick,->,color=black!80] (E).. controls (-1,0) and (-1,3 ) .. (A) ;
\draw[thick,->,color=black!80] (E).. controls (2,0) and (2,3 ) .. (B) ;

\node [shape=circle,fill=black!20,draw] (A) at     ($ ({9},{3}) $) {1};
\node [shape=circle,fill=black!20,draw] (B) at    ($ ({10},{3}) $) {2};
\node [shape=circle,fill=black!20,draw] (C) at    ($ ({9.5},{2}) $) {3};
\node [shape=circle,fill=black!20,draw] (D) at    ($ ({9.5},{1}) $) {4};
\node [shape=circle,fill=black!20,draw] (E) at    ($ ({9.5},{0}) $) {5};
\draw[thick,->,color=blue!80] (A) to (C) ;
\draw[thick,->,color=blue!80] (C) to (D) ;
\draw[thick,->,color=blue!80] (D) to (E) ;
\draw[thick,->,color=blue!80] (E) .. controls (11,0) and (11,3 ) .. (B) ;
\draw[thick,->,color=blue!80] (B) to (A) ;


\node [shape=circle,fill=black!20,draw] (A) at     ($ ({5},{3}) $) {1};
\node [shape=circle,fill=black!20,draw] (B) at    ($ ({6},{3}) $) {2};
\node [shape=circle,fill=black!20,draw] (C) at    ($ ({5.5},{2}) $) {3};
\node [shape=circle,fill=black!20,draw] (D) at    ($ ({5.5},{1}) $) {4};
\node [shape=circle,fill=black!20,draw] (E) at    ($ ({5.5},{0}) $) {5};
\draw[thick,->,color=blue!80] (A) to (C) ;
\draw[thick,->,color=blue!80] (C) to (B) ;
\draw[thick,->,color=blue!80] (B) .. controls (6,2.5) and (6,2 ) .. (D) ;
\draw[thick,->,color=blue!80] (E).. controls (4,0) and (4,3 ) .. (A) ;
\draw[thick,->,color=blue!80] (D) to (E) ;


\end{tikzpicture}
\caption{A strong  5 tournament (left, black) and the corresponding 2 Hamiltonian cycles (right, blue). }
\label{fig:tour5Ver-part}
\end{figure}
In this case there are two possible equivalence classes of configurations  that can be minimizers of the energy that are $13452$ (right Hamiltonian cycle in Figure  \ref{fig:tour5Ver-part}) and $13245$ (left Hamiltonian cycle in Figure  \ref{fig:tour5Ver-part}). Since it is possible to transform one equivalence class into the other by just two transposition we obtain
\begin{equation}\label{scomparsa}
H_5(13245)-H_5(13452)=\gamma(2,5)+\gamma(2,4)<0\,.
\end{equation}
The inequality follows by the fact that $(2,5)$ and $(2,4)$ are edges of the tournament so that both terms on the right hand side of \eqref{scomparsa} are negative. This means that there will be always one single minimizer that is 13245. Even if there are 2 Hamiltonian cycles the structure of the minimizers is like in the previous cases.
\begin{figure}
\begin{tikzpicture}[scale=0.9]
\node [shape=circle,fill=black!20,draw] (A) at      (0, 0) {1};
\node [shape=circle,fill=black!20,draw] (B) at      (0, -1) {2};
\node [shape=circle,fill=black!20,draw] (C) at      (-0.5, -2) {3};
\node [shape=circle,fill=black!20,draw] (D) at      (0.5, -2) {4};
\node [shape=circle,fill=black!20,draw] (E) at      (0,-3) {5};

\draw[thick,->,color=black!80] (B) to (C) ;
\draw[thick,->,color=black!80] (C) to (D) ;
\draw[thick,->,color=black!80] (D) to (B) ;
\draw[thick,->,color=black!80] (E).. controls (1.555,-3) and (1.555,0 ) .. (A) ;

\node [shape=circle,fill=black!20,draw] (A) at     ($ ({3.5},{0}) $) {1};
\node [shape=circle,fill=black!20,draw] (B) at    ($ ({3.5},{-1}) $) {2};
\node [shape=circle,fill=black!20,draw] (C) at    ($ ({3},{-2}) $) {3};
\node [shape=circle,fill=black!20,draw] (D) at    ($ ({4},{-2}) $) {4};
\node [shape=circle,fill=black!20,draw] (E) at    ($ ({3.5},{-3}) $) {5};
\draw[thick,->,color=blue!80] (A) to (B) ;
\draw[thick,->,color=blue!80] (B) to (C) ;
\draw[thick,->,color=blue!80] (C) to (D) ;
\draw[thick,->,color=blue!80] (D) to (E) ;
\draw[thick,->,color=blue!80] (E) .. controls (5,-3) and (5,-0 ) .. (A) ;


\node [shape=circle,fill=black!20,draw] (A) at     ($ ({6.5},{0}) $) {1};
\node [shape=circle,fill=black!20,draw] (B) at    ($ ({6.5},{-1}) $) {2};
\node [shape=circle,fill=black!20,draw] (C) at    ($ ({6},{-2}) $) {3};
\node [shape=circle,fill=black!20,draw] (D) at    ($ ({7},{-2}) $) {4};
\node [shape=circle,fill=black!20,draw] (E) at    ($ ({6.5},{-3}) $) {5};
\draw[thick,->,color=blue!80] (A) .. controls (7,-1) and (7, -1.5) .. (D) ;
\draw[thick,->,color=blue!80] (D) to (B) ;
\draw[thick,->,color=blue!80] (B) to (C) ;
\draw[thick,->,color=blue!80] (C) to (E) ;
\draw[thick,->,color=blue!80] (E) .. controls (8.055,-3) and (8.055,0 ) .. (A) ;

\node [shape=circle,fill=black!20,draw] (A) at     ($ ({9.5},{0}) $) {1};
\node [shape=circle,fill=black!20,draw] (B) at    ($ ({9.5},{-1}) $) {2};
\node [shape=circle,fill=black!20,draw] (C) at    ($ ({9},{-2}) $) {3};
\node [shape=circle,fill=black!20,draw] (D) at    ($ ({10},{-2}) $) {4};
\node [shape=circle,fill=black!20,draw] (E) at    ($ ({9.5},{-3}) $) {5};
\draw[thick,->,color=blue!80] (A)  .. controls (9, -1) and (9, -1.5) .. (C) ;
\draw[thick,->,color=blue!80] (C) to (D) ;
\draw[thick,->,color=blue!80] (D) to (B) ;
\draw[thick,->,color=blue!80] (B) to (E) ;
\draw[thick,->,color=blue!80] (E) .. controls (11,-3) and (11,0 ) .. (A) ;

\end{tikzpicture}
\caption{A strong  5 tournament (left, black) and the corresponding 3 Hamiltonian cycles (right, blue). }
\label{fig:tour5Ver-part-bis}
\end{figure}
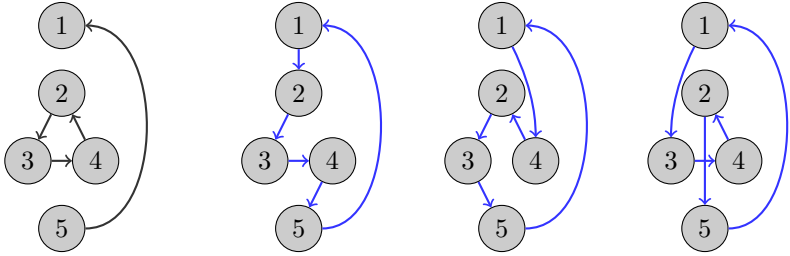

\smallskip

Let us consider now the case of the tournament of Figure 9. In this case there are 3 different Hamiltonian cycles that are
12345, 14235 and 13425. It is possible to transform the configuration 12345 into one of the other two with two transpositions and consequently we have
\begin{align*}
H_5(12345)-H_5(14235)&=\gamma(2,4)+\gamma(3,4)  \\
H_5(12345)-H_5(13425)&=\gamma(2,3)+\gamma(2,4)\,.
\end{align*}
In this case the difference between the energies is given by the sum of two terms one of which is positive while the other one is negative. Indeed it is not difficult to show that it is possible to fix the positive values of $\gamma(2,4), \gamma(4,3)$ and $\gamma(3,2)$ arbitrarily having always a choice of the remaining parameters in such a way that the tournament associated to the interaction is the one in Figure 9. This means that the ordering between the values of the energies of the possible ground states can be arbitrary. This means that we can have cases with one single minimizer that can be one between 12345, 14235 and 13425. We can have situations with two ground states with coinciding values of the energy. Finally In the special case $\gamma(2,4)=\gamma(4,3)=\gamma(3,2)$ we have that the 3 values of the energies are coinciding.

Consider for example the case $H_5(12345)=H_5(14235)<H_5(13245)$. Consider a model with $\frac{N}{5}$ particles for each type. This is a situation in which the configurations of particles minimizing the energy are not just the ones having all the particles of the same type grouped together in a single cluster. Let us call $C_j$ a generic cluster of particles of type $j$. With this notation the equivalence class of configurations $C_1C_2C_3C_4C_5$ is an equivalence class of particles in which going anticlockwise we meet before a cluster of particles of type 1 then of type 2 then of type 3, 4 and finally 5. This is necessarily a configuration having all the particles of the same type grouped together simply because we meet particle of a given type just once. The equivalence class  $C_1C_2C_3C_4C_5$ is a minimizer since all the  $\left(\frac N5\right)^{5}$ terms contributing to the energy in section \ref{ktypesec} obtained selecting one particle for each type will be equal to $H_5(12345)$
that is minimal. The same happens for the equivalence class $C_1C_4C_2C_3C_5$. Consider now an equivalence class of the form
$C_1C_4C_2C_3C_4C_5$. This is a class of configuration in which the particles of type $4$ can belong to two different clusters whose location is however determined by the above sequence. All the particles of the other types are joined together in single clusters. A configuration of this type is also a minimizer of the energy. Indeed when we select a particle for each type the only freedom that we have is in selecting a particle of type $4$ from one of the 2 clusters. Whatever cluster we choose we obtain that the five particles selected will be ordered like $12345$ or $14235$ and in both cases the contribution of the term will be minimal. Indeed these are all the minimizers because if we choose any other configuration of particles we will get at least a contribution that is not minimal.
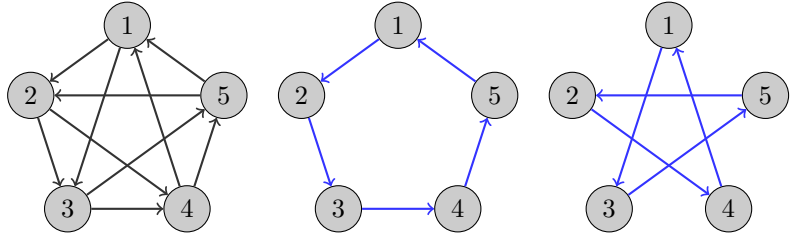
\begin{figure}
\begin{tikzpicture}[scale=0.9]
\node [shape=circle,fill=black!20,draw] (A) at     ($ ({1.5*cos(90)},{1.5*sin(90)}) $) {1};
\node [shape=circle,fill=black!20,draw] (B) at    ($ ({1.5*cos(90+72)},{1.5*sin(90+72)}) $) {2};
\node [shape=circle,fill=black!20,draw] (C) at    ($ ({1.5*cos(90+2*72)},{1.5*sin(90+2*72)}) $) {3};
\node [shape=circle,fill=black!20,draw] (D) at    ($ ({1.5*cos(90+3*72)},{1.5*sin(90+3*72)}) $) {4};
\node [shape=circle,fill=black!20,draw] (E) at    ($ ({1.5*cos(90+4*72)},{1.5*sin(90+4*72)}) $) {5};

\draw[thick,->,color=black!80] (A) to (B) ;
\draw[thick,->,color=black!80] (A) to (C) ;
\draw[thick,->,color=black!80] (B) to (D) ;
\draw[thick,->,color=black!80] (B) to (C) ;
\draw[thick,->,color=black!80] (C) to (D) ;
\draw[thick,->,color=black!80] (D) to (A) ;
\draw[thick,->,color=black!80] (D) to (E) ;
\draw[thick,->,color=black!80] (E) to (A) ;
\draw[thick,->,color=black!80] (E) to (B) ;
\draw[thick,->,color=black!80] (C) to (E) ;
\node [shape=circle,fill=black!20,draw] (A) at     ($ ({1.5*cos(90)+4},{1.5*sin(90)}) $) {1};
\node [shape=circle,fill=black!20,draw] (B) at    ($ ({1.5*cos(90+72)+4},{1.5*sin(90+72)}) $) {2};
\node [shape=circle,fill=black!20,draw] (C) at    ($ ({1.5*cos(90+2*72)+4},{1.5*sin(90+2*72)}) $) {3};
\node [shape=circle,fill=black!20,draw] (D) at    ($ ({1.5*cos(90+3*72)+4},{1.5*sin(90+3*72)}) $) {4};
\node [shape=circle,fill=black!20,draw] (E) at    ($ ({1.5*cos(90+4*72)+4},{1.5*sin(90+4*72)}) $) {5};
\draw[thick,->,color=blue!80] (A) to (B) ;
\draw[thick,->,color=blue!80] (B) to (C) ;
\draw[thick,->,color=blue!80] (C) to (D) ;
\draw[thick,->,color=blue!80] (D) to (E) ;
\draw[thick,->,color=blue!80] (E) to (A) ;

\node [shape=circle,fill=black!20,draw] (A) at     ($ ({1.5*cos(90)+8},{1.5*sin(90)}) $) {1};
\node [shape=circle,fill=black!20,draw] (B) at    ($ ({1.5*cos(90+72)+8},{1.5*sin(90+72)}) $) {2};
\node [shape=circle,fill=black!20,draw] (C) at    ($ ({1.5*cos(90+2*72)+8},{1.5*sin(90+2*72)}) $) {3};
\node [shape=circle,fill=black!20,draw] (D) at    ($ ({1.5*cos(90+3*72)+8},{1.5*sin(90+3*72)}) $) {4};
\node [shape=circle,fill=black!20,draw] (E) at    ($ ({1.5*cos(90+4*72)+8},{1.5*sin(90+4*72)}) $) {5};
\draw[thick,->,color=blue!80] (A) to (C) ;
\draw[thick,->,color=blue!80] (C) to (E) ;
\draw[thick,->,color=blue!80] (E) to (B) ;
\draw[thick,->,color=blue!80] (B) to (D) ;
\draw[thick,->,color=blue!80] (D) to (A) ;

\end{tikzpicture}
\caption{A strong  5 tournament (left, black) and the corresponding 2 Hamiltonian cycles (right, blue). }
\label{fig:tour5Ver-part-tris}
\end{figure}

\smallskip

Finally we consider the tournament of Figure 10 for which there are two different Hamiltonian cycles that are
12345 and 13524. In this case it is possible to transform one configuration into the other with 3 elementary transpositions and we obtain
\begin{equation}\label{uf}
H_5(12345)-H_5(13524)=\gamma(2,3)+\gamma(4,5)+\gamma(2,5)\,.
\end{equation}
Also in this case the terms on the right hand side of \eqref{uf} have different signs and both the Hamiltonian cycles can be associated to the true minimizer. It is possible to have also the situation in which the 2 values of the energies are coinciding. In this case however considering the model with $\frac N5$ particles of each type there are only two minimizers of the energy (up to translations). These are the configurations $C_1C_2C_3C_4C_5$ and $C_1C_3C_5C_2C_4$. It is not possible to have particles of the same type belonging to different clusters since otherwise  in the computation of the energy with \eqref{kbody}, \eqref{kaltra} there are some non minimal terms contributing. Essentially this follows by the fact that in the previous case (in the case of the 4 alphabet model) a minimizer can be obtained from the other just changing the position of one single particle while this is not the case in this example. The example briefly discussed in Section VI in \cite{EKKM2} should have an interaction of this type.

\smallskip

A classification of tournaments with $k=6$ is available (see the appendix of \cite{Moon}) but we are not going to give a detailed discussion of the models associated.

\section*{Acknowledgments} We thank an anonymous referee for a very careful reading of the paper and for the comments that improved the presentation of the paper.

\end{document}